\documentclass[namedreferences]{SolarPhysics}
\usepackage[optionalrh]{spr-sola-addons} 
\usepackage{graphicx}
\usepackage{url}             

\newcommand{\adv}{    {\it Adv. Space Res.}}

\newcommand{\aap}{    {\it Astron. Astrophys.}}

\newcommand{\ag}{     {\it Ann. Geophys.}}

\newcommand{\apj}{    {\it Astrophys. J.}}
\newcommand{\apjl}{   {\it Astrophys. J. Lett.}}

\newcommand{\grl}{    {\it Geophys. Res. Lett.}}

\newcommand{\jastp}{  {\it J. Atmos. Solar-Terr. Phys.}}
\newcommand{\jgr}{    {\it J. Geophys. Res.}}

\newcommand{\pasj}{   {\it Publ. Astron. Soc. Japan}}

\newcommand{\solphys}{{\it Solar Phys.}}

\newcommand{\ssr}{    {\it Space Sci. Rev.}}

\begin{document}
\begin{article}
\begin{opening}

\title{A Challenging Solar Eruptive Event of 18 November 2003
and the Causes of the 20 November Geomagnetic Superstorm. II.
CMEs, Shock Waves, and Drifting Radio Bursts}

\author{V.V.~\surname{Grechnev}$^{1}$\sep
        A.M.~\surname{Uralov}$^{1}$\sep
        I.M.~\surname{Chertok}$^{2}$\sep
        V.A.~\surname{Slemzin}$^{3}$\sep
        B.P.~\surname{Filippov}$^{2}$\sep
        Ya.I.~\surname{Egorov}$^{1}$\sep
        V.G.~\surname{Fainshtein}$^{1}$\sep
        A.N.~\surname{Afanasyev}$^{1}$\sep
        N.P.~\surname{Prestage}$^{4}$\sep
        M.~\surname{Temmer}$^{5}$}

\runningauthor{Grechnev et al.} \runningtitle{CMEs, Shock Waves,
and Drifting Radio Bursts in the 18 November 2003 Event}

\institute{$^{1}$ Institute of Solar-Terrestrial Physics SB RAS,
Lermontov St.\ 126A, Irkutsk 664033, Russia
                  email: \url{grechnev@iszf.irk.ru} \\ 
           $^{2}$ Pushkov Institute of Terrestrial Magnetism,
                  Ionosphere and Radio Wave Propagation (IZMIRAN),
                  Moscow 142190, Russia
                  email: \url{ichertok@izmiran.ru} \\ 
           $^{3}$ P.N. Lebedev Physical Institute, Leninsky Pr., 53, Moscow 119991,
                  Russia
                  email: \url{slem@lebedev.ru}\\
           $^{4}$ IPS Radio and Space Services Culgoora Solar Observatory, Narrabri, Australia
                  email: \url{nigel@ips.gov.au}\\
           $^{5}$ IGAM/Kanzelh{\"o}he Observatory, Institute of Physics, UNI Graz
                  email: \url{manuela.temmer@uni-graz.at}}

\date{Received ; accepted }

\begin{abstract}
We continue our study (Grechnev {\it et al.} (2013),
doi:10.1007/s11207-013-0316-6; Paper I) on the 18 November 2003
geoffective event. To understand possible impact on geospace of
coronal transients observed on that day, we investigated their
properties from solar near-surface manifestations in extreme
ultraviolet, LASCO white-light images, and dynamic radio spectra. We
reconcile near-surface activity with the expansion of coronal mass
ejections (CMEs) and determine their orientation relative to the
earthward direction. The kinematic measurements, dynamic radio
spectra, and microwave and X-ray light curves all contribute to the
overall picture of the complex event and confirm an additional
eruption at 08:07\,--\,08:20~UT close to the solar disk center
presumed in Paper~I. Unusual characteristics of the ejection appear
to match those expected for a source of the 20 November superstorm
but make its detection in LASCO images hopeless. On the other hand,
none of the CMEs observed by LASCO seem to be a promising candidate
for a source of the superstorm being able to produce, at most, a
glancing blow on the Earth's magnetosphere. Our analysis confirms
free propagation of shock waves revealed in the event and reconciles
their kinematics with ``EUV waves'' and dynamic radio spectra up to
decameters.
\end{abstract}

\end{opening}

\section{Introduction}

The geomagnetic storm on 20 November 2003 with Dst $= -422$~nT was
the strongest one after the destructive superstorm on 13\,--\,14
March 1989 (Dst $= -589$~nT) and has not been surpassed since. The
causes of the extreme nature of the 20 November 2003 superstorm and
its solar source remain unclear in spite of several attempts to
understand them (\textit{e.g.}, \opencite{Gopal2005};
\opencite{Yermolaev2005}; \opencite{Yurchyshyn2005};
\opencite{Ivanov2006}; \opencite{Moestl2008};
\opencite{Chandra2010}; \opencite{KumarManoUddin2011};
\opencite{Marubashi2012}; \opencite{Cerrato2012}, and others). The
challenge of this superstorm urged us to investigate into various
aspects of the 18 November solar eruptive event in active region
(AR) 10501 that is considered to be its only possible source.
Eruptions from AR~10501 have been addressed in Paper~I
\cite{Grechnev2013_I}. Its conclusions are: (i)~eruption at 07:29
(all times are referred to UT) produced a missed M1.2 flare probably
associated with onset of the first southeast coronal mass ejection,
CME1; (ii)~eruptions before 07:55 are unlikely to be responsible for
the superstorm; (iii)~the eruptive filament collided with a
topological discontinuity, bifurcated, and transformed into a
Y-shaped cloud, which had not left the Sun; thus, the filament
should not be directly related to the magnetic cloud hitting the
Earth; (iv)~one more eruptive episode possibly occurred between
08:07 and 08:17 that could be related to the disintegration of the
filament and led to other consequences open to question.

All of the listed studies assumed that the source of the superstorm
was either the southeast CME1 observed by the \textit{Large Angle
and Spectroscopic Coronagraph} (LASCO; \opencite{Brueckner1995})
starting from 08:06 or, more probably, the second southwest halo CME
(CME2), which appeared at 08:49. According to the model of the cone
CME geometry (\textit{e.g.}, \opencite{Howard1982};
\opencite{FisherMunro1984}), the halo shape indicates the earthward
(or the opposite) propagation of a CME. Therefore, CME2 has been
considered as the major candidate for the source of the superstorm.
On the other hand, it is possible that the outer halo of CME2 was a
trace of a shock front. If so, then CME2 was not necessarily
Earth-directed. Thus, it is necessary to find out the nature of the
structural components of CME2 and its actual orientation.

One more challenge of this event is the mismatch between the
right-handed helical magnetic cloud (MC) and the pre-eruption region
of left-handed helicity established by \inlinecite{Chandra2010}. To
resolve the problem, the authors proposed a right-handed helical
ejection from a minor area of AR~10501. Based on this idea,
\inlinecite{KumarManoUddin2011} related CME1 to a partial eruption
at 07:41 from this area and proposed a merger of the magnetic
structures of CME1 (presumably right-handed) and CME2. The authors
supported the interaction between the CMEs by a drifting radio burst
observed by \textit{Wind}/WAVES around 09:00. Another attempt to
understand the encounter of the MC with the Earth based on the
conjecture of \inlinecite{Chandra2010} was made by
\inlinecite{Marubashi2012} who considered that the MC evolved from a
single right-handed CME. Neither of these studies presented a
quantitative confirmation of their conjectures, whilst attributing
the superstorm to a partial eruption from a minor region seems to be
questionable.

Paper~I concluded that CME1 was probably initiated in the east,
excessively left-handed, part of AR~10501 at 07:29 (consistent with
an estimate of \opencite{Gopal2005}) in association with an
unreported M1.2 flare thus contradicting the interpretations of
\inlinecite{Chandra2010}, \inlinecite{KumarManoUddin2011}, and
\inlinecite{Marubashi2012}. This is why the source region of CME1 is
important.

The present paper (Paper~II) is focused on CME1 and CME2 and the
probable nature of their components. In order to understand their
possible geoeffective implications, we in particular address the
following questions: when and where was CME1 initiated, how was CME2
directed with respect to the Earth, and what erupted between 08:07
and 08:17 close to the solar disk center. We specify measurements of
\inlinecite{Gopal2005} and confirm the results by comparing them to
signatures of shock waves in dynamic radio spectra at metric and
decametric wavelengths as well as their possible near-surface
traces. White studying this particular event, we pursue a better
understanding of CMEs and related phenomena.

Section~\ref{S-kinematics} describes our measurement techniques.
Section~\ref{S-overview} outlines the pre-event situation and its
overall evolution. Section~\ref{S-observations} analyzes the
observations. The results are discussed in
Section~\ref{S-discussion} and summarized in
Section~\ref{S-conclusion}.

\section{Measurement Techniques}
 \label{S-kinematics}

Two kinds of transients appear in LASCO images: magnetoplasma CME
components (henceforth `mass ejections' or `CMEs') and traces of
waves (\opencite{Sheeley2000}; \opencite{Vourlidas2003};
\citeauthor{Grechnev2011_I},
\citeyear{Grechnev2011_I,Grechnev2011_III}). The kinematics of the
two kinds of transient are different. This section describes
kinematics of non-wave and wavelike transients and methods of
measurement.

We consider two kinds of wave signatures in LASCO images: faint
non-structured (or structured by coronal rays) halo-like outermost
envelopes of CMEs and deflections of coronal streamers. The
brightness of the halos can be very low. Mass ejections are
significantly brighter, with well pronounced loops or threads in
their structure. It is difficult to reliably identify both wave
signatures and CME structures in a single set of images. We
therefore use two separate sets processed in different ways to
measure wave traces and mass ejections. For CMEs we use ratios of
current LASCO images $C(j)$ to a fixed pre-event image $C(0)$ and
limit the values in the ratios from both above and below with
thresholds $A_0 \lsim 1$ and $A_1 \gsim 1$, $A_0 < I_\mathrm{CME}(j)
= C(j)/C(0) < A_1$. For wave signatures we use ratios of running
differences $C(j)-C(j-1)$ to preceding images $C(j-1)$ also with
optimized contrast by adjusting the corresponding thresholds $B_0
\lsim 0$ and $B_1 \gsim 0$, $B_0 < I_\mathrm{wave}(j) =
[C(j)-C(j-1)]/C(j-1) < B_1$.

\subsection{Mass Ejections}
 \label{S-cme_expansion}

The kinematics of coronal transients have been measured in several
different ways. Height-time plots are obtained by measuring a
characteristic CME feature. Then the measurements are differentiated
(\textit{e.g.}, \opencite{Maricic2004}; \citeauthor{Temmer2008},
\citeyear{Temmer2008, Temmer2010}). Alternatively, the measurements
are fit with an analytic function such as polynomial
\cite{Yashiro2004,Gopal2009}, Gaussian \cite{WangZhangShen2009}, or
more sophisticated models \cite{KrallChenSantoro2000}.

Both approaches should converge to similar results, but each method
has its shortcomings. Differentiation of measurements is critical to
temporal sampling, errors, and provides large uncertainties. The
adequacy of an analytic fit might be questionable. For example, the
polynomial fit used in the SOHO LASCO CME Catalog
(\opencite{Yashiro2004}; \opencite{Gopal2009},
\url{http://cdaw.gsfc.nasa.gov/CME_list/}) is probably the best way
for approximately evaluating the kinematics of CMEs, but the
underlying assumption of a constant (or zero) acceleration
(\textit{i.e.}, the constancy of the driving/retarding force) does
not seem to be realistic. Employment of theoretical models like the
flux rope model of \citeauthor{Chen1989} (\citeyear{Chen1989,
Chen1996}; \textit{e.g.}, \opencite{KrallChenSantoro2000}) is
complex, whereas its veracity has not been established.

Our way is based on self-similarity of CME expansion (see,
\textit{e.g.}, \opencite{Illing1984};
\opencite{CremadesBothmer2004}). The theory of self-similar
expansion of solar CMEs was developed by \inlinecite{Low1982}. A
description of a self-similar expansion convenient for analysis of
observations was proposed by \inlinecite{Uralov2005}. A self-similar
expansion of an individual plasma packet under the frozen-field
conditions and negligible drag of the medium is described by an
equation
\begin{equation}
\rho \frac{d{\bf v}}{dt}=\frac 1{4\pi }{\bf {rotB\times B}}-{\bf
{grad}}{p}- \rho \frac{GM_{\odot}}{r^2}{\bf {e}_{\bf {r}}} = \\
{\bf {F}}_B+{\bf {F}}_p+{\bf {F}}_g,
 \label{E-momentum}
\end{equation}
where $p$ and $\rho$ are the gas pressure and density; ${\bf B}$ the
magnetic field vector, ${\bf v}$ the velocity, $M_{\odot}$ the mass
of the Sun, and $G$ the gravitational constant. ${\bf F}_B$, ${\bf
F}_p$, and ${\bf F}_g$ are the total magnetic, plasma pressure, and
gravitational forces affecting the unit volume. Let $R=R(t)$ be some
spatial scale characterizing the size of the expanding region at the
instant $t$. The forces in Equation (\ref{E-momentum}) depend on the
distance $R$ as
\begin{eqnarray}
|{\bf F}_B| \propto \left( \frac{R_0}{R}\right)^4 \frac{1}{R}, \quad
|{\bf F}_p| \propto \left( \frac{R_0}{R}\right)^{3\gamma}
\frac{1}{R}, \quad |{\bf F}_g| \propto \left( \frac{R_0}{R}\right)^3
\frac{1}{R^2},
 \label{E-f_dependence}
\end{eqnarray}
where $R_0$ is the initial size of the self-similar expansion. Force
${\bf F}_B$ combines all magnetic forces affecting the expanding
packet including propelling magnetic pressure and retarding magnetic
tension. Force ${\bf F}_p$ due to plasma pressure is directed
outward. The gravitational force ${\bf F}_g$ retards expansion. With
a polytropic index $\gamma =4/3$, all the terms in Equation
(\ref{E-f_dependence}) which appear in the right-hand-side of
Equation (\ref{E-momentum}) decrease synchronously with distance and
time by the same scaling factor preserving orientation. This fact
determines the self-similar expansion of the ejecta. From the
expressions of \inlinecite{Uralov2005}, the instant velocity $v$ can
be related to the distance from the expansion center $R$
\cite{Grechnev2008}:
\begin{eqnarray}
v^2 = v_0^2+\left(v_\infty^2 - v_0^2\right)\left({1-R_0/R}\right),
 \label{E-expansion_vel}
\end{eqnarray}
where $v = dR/dt$ and $v_0$ and $v_\infty$ are the initial and
asymptotic velocities of the self-similar expansion stage. Analysis
of this expression shows the following \cite{Grechnev2011_I}.

\begin{enumerate}
\item
 Acceleration of the ejecta in self-similar expansion can only decrease by
the absolute value or be exactly zero. Therefore, \textit{the
self-similar approach does not apply to initial stages, when the
acceleration increases.}

\item
 Acceleration $a$, if nonzero, goes at large distances ($R \approx r \gg
R_\odot$) as $|a| \propto r^{-2} \to 0$. Thus,
\textit{self-similar expansion cannot be fit with any polynomial.}

\item
 Three expansion regimes are possible:

(a)~accelerating ejecta, $v_0 < v_\infty$;

(b)~decelerating ejecta, $v_0 > v_\infty$ (`explosive' eruption);

(c)~inertial expansion, $v_0 = v_\infty$.

\end{enumerate}

The accelerating regime (a) probably applies to all
non-flare-related CMEs and many flare-related ones. In cases (b)
and (c), a strong initial impulsive acceleration occurs before the
onset of the self-similar stage.

Integrating Equation (\ref{E-expansion_vel}), despite its
simplicity, cannot provide an explicit distance \textit{vs.} time
dependence. The following expression allows one to calculate a
self-similar expansion implicitly, as time $t$ \textit{vs.} the
heliocentric distance $r$, given the distance of the eruption center
$r_\mathrm{c}$ and the CME velocity $v_1$ measured at time $t_1$ at
a distance $r_1$:
\begin{eqnarray}
 t(r) = t_1 + 1/v_\infty^3 \times
 \qquad \qquad \qquad \nonumber\\
 \left\{
    S v_\infty \sqrt{r-r_\mathrm{c}} - v_\infty v_1 r_1
                   + (v_\infty^2-v_1^2) r_1 \ln \left[ \frac{ v_\infty \sqrt{r-r_\mathrm{c}} + S } {(v_\infty +
                    v_1) \sqrt{r_1}}\right]
                    \right\}
                    \label{E-self_sim_exp} \\
 \mathrm{with} \quad S = \sqrt{v_\infty^2 (r-r_\mathrm{c}-r_1)+v_1^2 r_1}. \nonumber
\end{eqnarray}
The initial estimates of $v_1$ and $v_\infty$ can be taken from the
CME catalog and improved iteratively. The onset time $t_0$ of a
self-similar expansion is:
\begin{eqnarray}
t_0 = \left\{
 \begin{array}{cc}
 t(r_\mathrm{c}) \;\; & \mathrm{for} \quad v_1 > v_\infty, \\
 t\left( \left[ {r_\mathrm{c} + r_1 \left( 1-
\frac{v_1^2}{v_\infty^2} \right) }
 \right]
 \right)  \;\; & \mathrm{for} \quad v_1 < v_\infty.
 \end{array}
 \right.
  \label{E-onset_time}
\end{eqnarray}

Monotonically decreasing or zero acceleration is consistent with
observations (see, \textit{e.g.}, \opencite{Zhang2001};
\opencite{ZhangDere2006}; \citeauthor{Temmer2008},
\citeyear{Temmer2008,Temmer2010}). Although the self-similar
approximation does not apply to the initial impulsive acceleration
stage, it promises a better fit to the observed CME expansion and
higher accuracy of the estimated onset time than the polynomial fit
does.

In specifying the CME onset times we also employ the temporal
closeness of the major CME acceleration with hard X-ray (HXR) or
microwave bursts revealed in the mentioned series of the papers as
well as the Neupert effect \cite{Neupert1968}. These circumstances
indicate that the CME velocity profile is roughly reflected in the
rising phase of the corresponding soft X-ray light curve recorded
with GOES.

\subsection{Waves}
 \label{S-wave_expansion}

CME-associated waves are most likely excited by abrupt eruptions of
magnetic ropes inside developing CMEs during rising hard X-ray and
microwave bursts \cite{Grechnev2011_I}. The waves rapidly steepen
into shocks, pass through the forming CME frontal structures, and
freely propagate afterwards for some time like decelerating blast
waves (\textit{cf.} \opencite{PomoellVainioKissmann2008}). The
corresponding quantitative description allows one to reconcile
manifestations of shocks in different emissions including Moreton
waves, `EUV waves', metric type II bursts, and leading edges of
CMEs. A narrowband harmonic type II burst appears if the shock front
compresses the current sheet of a coronal streamer, producing a
running flare-like process \cite{Uralova1994}.

A simple model (\citeauthor{Grechnev2008}
\citeyear{Grechnev2008,Grechnev2011_I,Grechnev2011_III}) describes
propagation of such a blast-like shock wave in plasma with a radial
power-law density falloff $\delta$ from an eruption center, $n =
n_0(x/h_0)^{-\delta}$. Here $x$ is the distance and $n_0$ is the
density at a distance of $h_0 \approx 100$ Mm, which is close to the
scale height. The propagation of a shock wave in the self-similar
approximation is determined by plasma density distribution, being
almost insensitive to the magnetic fields. Such a wave decelerates
if $\delta < 3$, due to a growing mass of swept-up plasma.
Propagation of such a shock \textit{vs.} time $t$ is described by an
expression $x(t) \propto t^{2/(5-\delta)}$, which is more convenient
for use in a form
\begin{eqnarray}
x(t) = x_1[(t-t_0)/(t-t_1)]^{2/(5-\delta)},
 \label{E-pl_fit}
\end{eqnarray}
where $t$ and $x$ are the current time and distance, $t_0$ is the
wave onset time, and $t_1$ and $x_1$ correspond to one of the
measured fronts.

To fit the drift of a type II burst, we take an initial estimate of
$\delta$ (typically $2 \leq \delta \leq 2.8$) and choose a reference
point on a band with a harmonic number $N_\mathrm{ref}$ (1 or 2) at
a frequency $f_\mathrm{ref}$ and time $t_1$. The corresponding
plasma density is $n_1 = [f_\mathrm{ref}(t_1) N_\mathrm{ref}^{-1}/(9
\times 10^3 )]^{2}$, and the height is $x_1 =
h_0\,(n_0/n_1)^{1/\delta}$. Then the height\,--\,time plot of the
shock tracer is calculated from Equation (\ref{E-pl_fit}); the
corresponding density variation is $ n(t) =
n_0\,[x(t)/h_0]^{-\delta}$. The trajectory of the
fundamental-emission type II band is $f_\mathrm{fund}(t) = 9 \times
10^3 [n(t)]^{1/2}$, and the trajectory of the harmonic-emission band
is $f_\mathrm{harm}(t) = 2 f_\mathrm{fund}(t)$. By adjusting
$\delta$ and $t_0$ in sequential attempts, we approach a best
trajectory of the bands \cite{Grechnev2011_I}. The spectrum can be
reconciled with measured heights by adjusting $n_0$, as usually
done.

Presumed traces of shocks in coronagraph images are fitted
similarly. Input parameters are starting estimates of $\delta$ and
$t_0$, the heliocentric distances of the wave origin $r_0$ and the
wave front $r_1$ measured at a time $t_1$. The initial
approximation of the height\,--\,time plot is
$r(t)=(r_1-r_0)\left[(t-t_0)/(t_1-t_0)\right]^{2/(5-\delta)} +
r_0$. Then a best fit is achieved in sequential attempts
(\citeauthor{Grechnev2011_I}
\citeyear{Grechnev2011_I,Grechnev2011_III}).

\subsection{Resizing Representation}
 \label{S-resize}

CMEs are usually analyzed by using images in which the spatial
resolution is fixed so that the Sun has the same size, while a CME
expands. Self-similarity of CME expansion can be used to improve
the accuracy of measurements. We adjust the spatial scale to fix
the CME size. This way reveals properties of CME expansion that
are difficult to notice in the usual representation.

We resize images according to a corresponding fit described in the
preceding sections to compensate expansion of a transient and keep
its visible size unchanged. In each of the resized images we outline
the whole transient with an oval by changing its parameters
according to an analytic fit and endeavor to catch the outer
contour. Fitting the whole transient rather than single feature
considerably improves the accuracy, and resizing all of the images
by a single fit allows us to neglect minor irregular deviations
between sequential images. Small systematic trends can be detected
and compensated for in looking at a movie composed from resized
images. Measurement accuracy can be farther improved in this way.

The resizing representation also (i)~facilitates detection of
deviations in expansion of CME components from a self-similar one
providing indications of their nature and revealing internal motions
in a CME, (ii)~allows measurements from CME flanks when its leading
edge departs from the field of view; (iii)~simplifies identification
of CME components visible in white light with structures observed in
different emissions at earlier stages of an eruption.

From the kinematics of CMEs and shock waves it follows that a CME
asymptotically approaches a fixed velocity, while a related shock
wave continuously decelerates. The relative distance between a fast
CME and the shock front decreases so that eventually it enters the
bow-shock regime. This probably occurs beyond the field of view of
LASCO-C3, while the approach of a CME to the leading wave front is
sometimes visible in resized images. If a CME is not fast enough,
then the shock decays to a weak fast-mode disturbance.

\section{Overview of the Event}
 \label{S-overview}

\subsection{Pre-event Situation}

\begin{figure} 
  \centerline{\includegraphics[width=\textwidth]
   {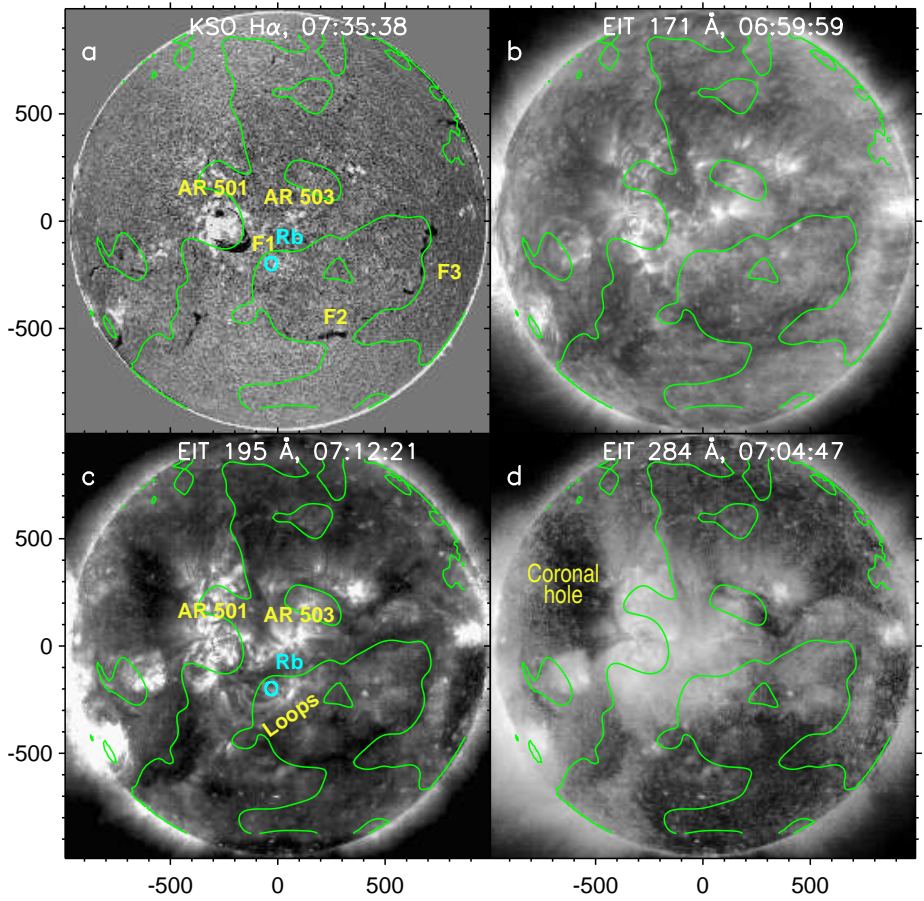}
  }
  \caption{Pre-event situation in a KSO H$\alpha$
image (a) and EIT images at 171~\AA\ (b), 195~\AA\ (c), and 284~\AA\
(d). The green contours present the magnetic neutral line. F1 is the
pre-eruption main filament, F2 an F3 are remote filaments. The
light-blue oval marks region Rb where the eruptive filament
bifurcated. The axes show the coordinates in arcsec from the solar
disk center.}
  \label{F-pre-event}
  \end{figure}

The pre-event situation is presented in Figure~\ref{F-pre-event}.
The H$\alpha$ image in Figure~\ref{F-pre-event}a (Kanzelh{\"o}he
Solar Observatory, KSO) shows a large U-shaped filament F1 rooted in
AR~10501 and pointed southwest. The pre-eruption filament was
inclined to the solar surface by $\approx 60^{\circ}$ ($\approx
23^{\circ}$ to the line of sight, see Paper~I). The green contours
show the neutral line of the line-of-sight magnetic component
($B_l$) at the photospheric level. The green contours are rather
coarse tracers mainly corresponding to dark filaments F1, F2, and F3
in the H$\alpha$ image, but deviating considerably from a
high-latitude southeast filament.

Southwest neighbors of AR~10501 were AR~10503 and region `Rb' (small
light-blue oval) where eruptive filament F1 bifurcated. Long loops
labeled in Figure~\ref{F-pre-event}c south from region Rb connected
a western plage region with the south edge of AR~10501.
Figures~\ref{F-pre-event}b and \ref{F-pre-event}c show that
filaments F2 and F3 visible in Figure~\ref{F-pre-event}a were
arranged along an extended channel still farther southwest. The
propagation of shock waves excited by eruptions could be affected by
density inhomogeneities indicated by brighter regions by the sides
of the filaments as well as a large coronal hole northeast of
AR~10501 in Figure~\ref{F-pre-event}d (EIT 284~\AA;
\opencite{Delab1995}).

\subsection{Time Profiles and Episodes of the Whole Event}

\begin{figure} 
  \centerline{\includegraphics[width=\textwidth]
   {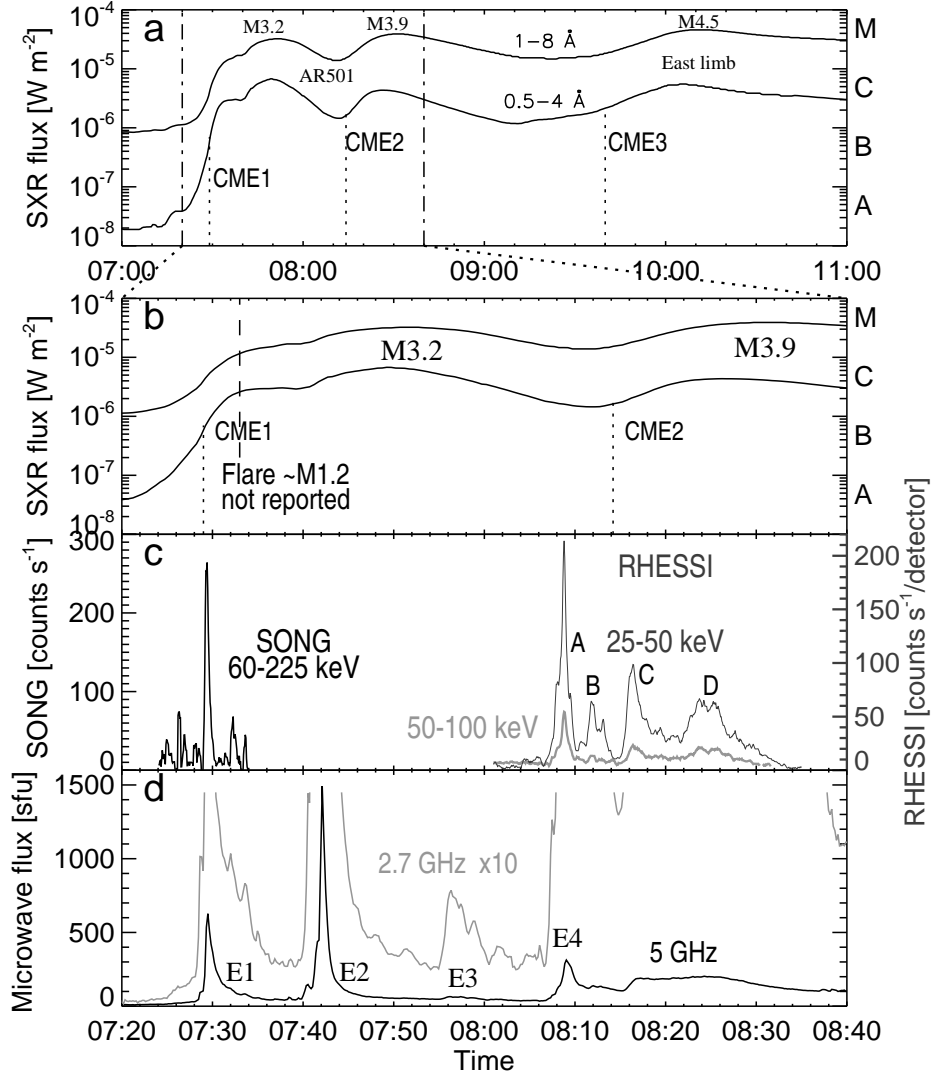}
  }
  \caption{Flare emissions throughout the whole event.
(a)~GOES SXR flux; (b)~its extended part in the interval marked with
dash-dotted lines in panel (a); (c)~hard X-ray flux; (d)~microwaves
at 5 GHz (black) and 2.7 GHz (gray, magnified by a factor of 10).}
  \label{F-timeprof}
  \end{figure}

Figure~\ref{F-timeprof} presents time profiles of soft (a,\,b) and
hard (c) X-ray emissions as well as microwaves (d) for the whole
event. The GOES soft X-ray (SXR) light curves are supplied with
comments on their importance, positions of the flares, and onset
times of the CMEs estimated by \inlinecite{Gopal2005} and specified
below. A detailed description is given in Paper~I.

Table~\ref{T-table1} lists associations of the flare peaks
E1\,--\,E4 with eruptive episodes according to Paper~I. Episode E1
with strong impulsive HXR and microwave bursts increased the SXR
flux up to $\approx\,$M1.2 level but was not reported as a separate
event. After E1, H$\alpha$ flare ribbons, a flare arcade, and EUV
dimmings have appeared. This episode is a candidate for the onset of
CME1, but a related eruption was not observed (TRACE had a gap in
observations). This caused confusion about the onset time of CME1 in
some preceding studies.

 \begin{table}
 \caption{Episodes of eruption in AR~10501 revealed in Paper~I.}
\label{T-table1}
 \begin{tabular}{lcl}
 \hline
\multicolumn{1}{c}{Episode} & \multicolumn{1}{c}{Peak time} & \multicolumn{1}{c}{Manifestation} \\
 \hline
E1 & 07:29  & Eruption in the east part of AR~10501. Unreported M1.2 flare \\
E2 & 07:41  & Impulsive jet-like ejection. Main filament F1 departs \\
E3 & 07:56  & Main filament F1 accelerates \\
E4A & 08:09  & Eruptive filament F1 collides with region of bifurcation \\
E4B & 08:12  & Eruptive filament F1 bifurcates \\
E4C & 08:16  & Region of bifurcation dims and disconnects from AR~10501 \\
E4D & 08:24  & Last flare episode (not considered in Paper I) \\
-- & 08:23\,--\,09:55  & Remnants of filament F1 move toward the limb as Y-like cloud \\

 \hline
 \end{tabular}

 \end{table}

An impulsive jetlike ejection erupted at 07:41 (E2) along the
southeast leg of filament F1 and then moved along the loops denoted
in Figure~\ref{F-pre-event}c. Paper~I concluded that development of
a CME in episode E2 was unlikely, but the sharp ejection could have
produced a shock. The latter conjecture is supported by a type II
burst, which was reported by several observatories starting from
07:47.

Filament F1 slowly departed after episode E2 and additionally
accelerated to 110~km~s$^{-1}$ during a weak episode E3. At about
08:07 the eruptive filament collided with region Rb and bifurcated.
The collision and subsequent phenomena were manifested in a
four-component flare observed in the H$\alpha$ line in KSO, in EUV
with TRACE, and in HXR with RHESSI (\citeauthor{Miklenic2007}
\citeyear{Miklenic2007,Miklenic2009}; \opencite{Moestl2008}). The
HXR peaks E4A and E4B (Figure~\ref{F-timeprof}c) had a response in
the bifurcation region Rb, indicating its connection with the flare
site in AR~10501 that later disappeared. Dimming developed in region
Rb at that time.

Then the bifurcated filament inverted and transformed into a large
dark Y-shaped cloud visible in the CORONAS-F/SPIRIT 304~\AA\ images
to move during 08:23\,--\,09:55 southwest toward the limb. The
fastest part of the Y-darkening had a speed of $\approx
210$~km~s$^{-1}$, and its main body which had an initial speed of
110~km~s$^{-1}$ decelerated, suggesting an almost constant real
speed nearly along the solar surface.

The transformation of the eruptive filament and disconnection of the
bifurcation region Rb from AR~10501 suggest one more significant
eruption during episodes E4A\,--\,E4C. Paper~III (Uralov \textit{et
al.}, in preparation) will consider what occurred in this region at
that time. A later eruption associated with an M4.5 SXR peak at
10:11 (Figure~\ref{F-timeprof}a) occurred at the east limb in a
rising region 10508 (return of AR~10486). Most likely, this event
was related to the third, large CME, whose extrapolated onset time
was about 09:40 \cite{Gopal2005}.

\subsection{CMEs}

\begin{figure} 
  \centerline{\includegraphics[width=\textwidth]
   {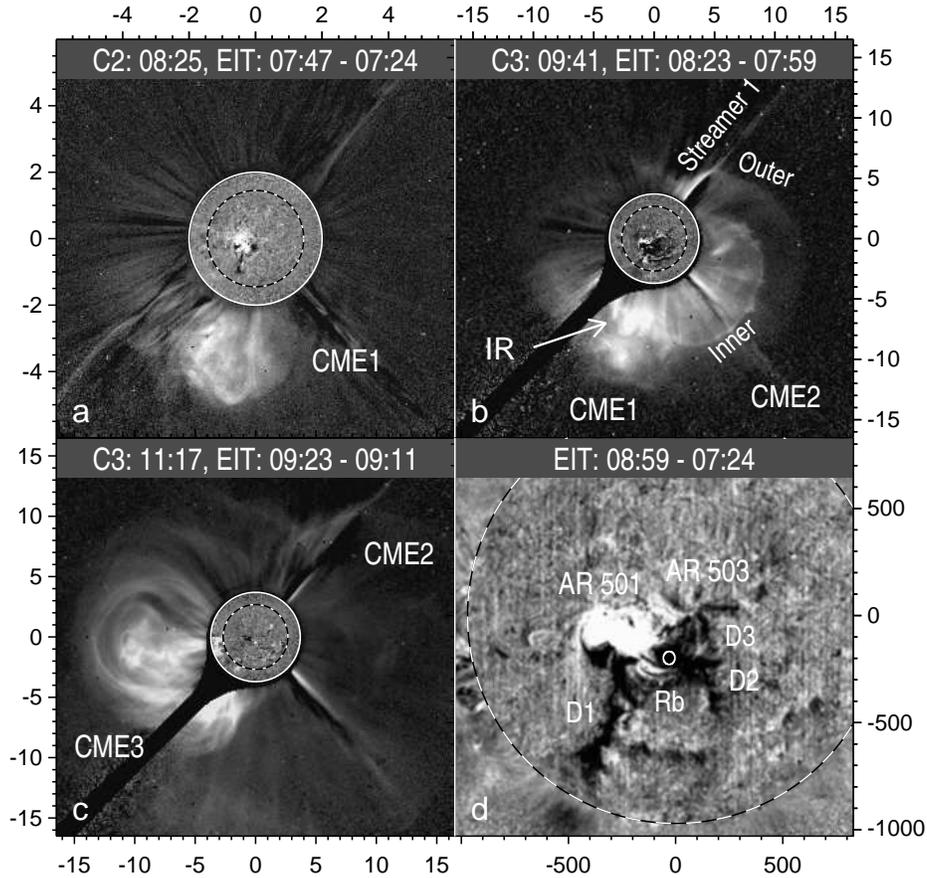}
  }
  \caption{Three major CMEs of 18 November (a\,--\,c) and
summary of surface activity in the post-event/pre-event EIT 195~\AA\
ratio image (d). EIT ratio images inserted into the LASCO ratio
images are magnified by factors 1.45 (a) and 2.68 (b,c) for better
viewing. The solid circles denote the inner boundaries of the fields
of view of the coronagraphs. The broken circles denote the solar
limb in EIT images. The axes show distances from the solar disk
center in solar radii (a\,--\,c) and in arcsec (d).}
  \label{F-3cmes}
  \end{figure}

Figures~\ref{F-3cmes}a and \ref{F-3cmes}c show LASCO ratio images of
three significant CMEs observed on that day
(\opencite{ChertokGrechnev2005}; \opencite{Grechnev2005};
\opencite{Gopal2005}). The EIT 195~\AA\ ratio images in the central
insets are magnified by factors 1.45 (a) and 2.68 (b,c) to better
show related surface activity. The EIT 195~\AA\ ratio image in
Figure~\ref{F-3cmes}d presents changes throughout the whole event in
AR~10501. CME1 and CME2 were related to this event.

The southeast CME1 (Figure~\ref{F-3cmes}a) appeared at 08:06. Its
linear-fit speed was 1223 km~s$^{-1}$. The onset time estimated by
\inlinecite{Gopal2005} was about 07:22. The volume of CME1 appears
to be filled with enhanced-density diffuse material and loop-like
structures. The CME structure approximately corresponds on the Sun
to an elongated south dimming, a deeper central dimming adjacent to
a bright arcade, and the arcade itself. The dimming and flare arcade
started to develop before 07:35 according to EUV and H$\alpha$ data
(Figure~4 of Paper~I) suggesting that the onset time of CME1 was
still earlier, most likely, corresponding to the flare episode E1 at
07:29. In addition to the relatively narrow south CME1, its faint
partial-halo extension is detectable in the whole eastern half of
the image suggesting an expanding wave disturbance.

The brighter, wider and faster southwest CME2
(Figure~\ref{F-3cmes}b) appeared at 08:49. Its east flank intruded
into CME1 (the intrusion region IR in the figure). The linear-fit
speed of the fastest feature of CME2 was 1660 km~s$^{-1}$.
\inlinecite{Gopal2005} detected the inner and outer components of
CME2 and estimated their onset times of about 08:08 and 08:20,
respectively. The structure of CME2 looks different from a
three-part one: neither a bright core nor dark cavity separating it
from the frontal structure were pronounced. The inner component
consisted of radial threadlike features, suggesting that it was an
expanding arcade. The faint outer halo component had a diffuse
non-structured body and a pronounced leading edge. This halo edge
crossed a distorted streamer~1 in Figure~\ref{F-3cmes}b well ahead
of the inner structure, suggesting an expanding shock wave
\cite{Sheeley2000,Vourlidas2003,Grechnev2011_I}. A large central
dimming in regions Rb and AR~10503 in Figure~\ref{F-3cmes}b suggests
location of a CME source region there.

A large southeast CME3 observed starting from 09:50
(Figure~\ref{F-3cmes}c) was not related to AR~10501
(\opencite{ChertokGrechnev2005}; \opencite{Grechnev2005};
\opencite{Gopal2005}). Most likely, CME3 was due to an eruption at
the east limb from a rising AR~10508 (former 10486), as an EIT image
in the inset shows, and corresponded to an M4.5 SXR flare, which
peaked at 10:11 (Figure~\ref{F-timeprof}a). The three-part structure
of CME3 was preceded by a fast faint halo (the average speed of 1824
km~s$^{-1}$), which deflected the streamers suggesting one more
shock wave. Magnetic structures of CME3 are not expected to have
reached the Earth, as preceding studies concluded. The only possible
implication of CME3 could be a lateral pressure from the associated
shock front to constrain expansion of the magnetic cloud responsible
for the 20 November superstorm.

The EIT 195~\AA\ ratio image in Figure~\ref{F-3cmes}d shows a bright
arcade in AR10501 (which looks saturated, because we show a narrow
range of the brightness) and dimmed regions. Dimming D1 developed in
association with CME1. Dimming D2 discussed in Paper~I developed
around region Rb, where the U-shaped filament bifurcated. A
star-like dimming D3 also appeared in region 10503 thus indicating
its involvement.

\section{Coronal Transients Observed During the Event}
 \label{S-observations}

\subsection{CME1 (08:05) and Wave 1}

\begin{figure} 
  \centerline{\includegraphics[width=0.6\textwidth]
   {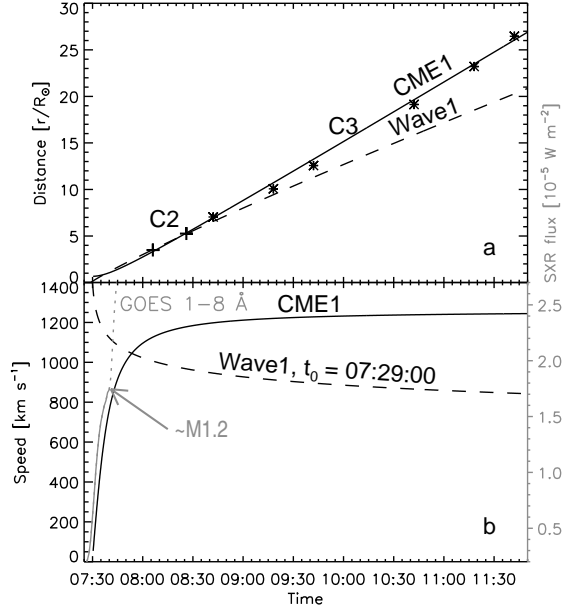}
  }
  \caption{Kinematical plots of CME1 (solid) and associated
wave~1 (dashed) visible in LASCO images in Figures
\ref{F-lasco_wave1} and \ref{F-lasco_cme1}. The symbols in panel
(a) present the measurements from the CME catalog. The dotted line
in panel (b) presents the GOES SXR flux at 1\,--\,8~\AA.}
  \label{F-cme1_plot}
  \end{figure}

A wide, faint halo-like extension of CME1 suggestive of an expanding
wave front is called hereafter wave~1. We fit the observed expansion
of the halo by using Equation (\ref{E-pl_fit}) from
Section~\ref{S-wave_expansion} and expansion of the CME1 main
structure by using Equations (\ref{E-self_sim_exp}) and
(\ref{E-onset_time}) from Section~\ref{S-cme_expansion}. The
measurement accuracy cannot be high because of the absence of
observations of a related eruption, and therefore we limit our
attempts by acceptable correspondence with available data. We use a
simpler accelerating kinematics, because it is not possible to
recognize whether CME1 accelerated or decelerated at large
distances. We also employ the mentioned expectation of similarity
between the rising parts of the SXR flux and the CME speed. The
kinematical plots are shown in Figure~\ref{F-cme1_plot}. The plots
for both CME1 and wave~1 converge to event E1 ($\approx \,$M1.2) at
about 07:29. A sharper rise of the SXR emission after 07:34 (the
dotted part of the GOES light curve) is due to the next episode E2.
The height-time plot of CME1 is close to the measurements in the CME
catalog denoted by symbols.

\begin{figure} 
  \centerline{\includegraphics[width=\textwidth]
   {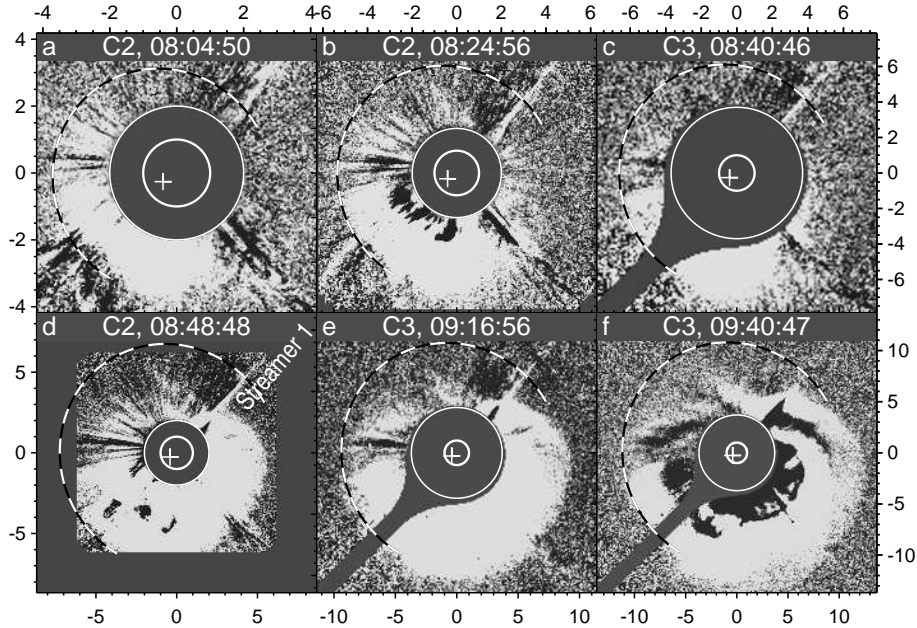}
  }
  \caption{The wave associated with CME1 in LASCO-C2 and C3
running-difference ratio images resized to compensate for the
expansion of the wave front. The dashed oval outlines the outermost
traces of wave~1. The circles denote the solar limb and the inner
boundaries of the fields of view of the coronagraphs. The cross
denotes the initial wave center. The axes show hereafter distances
from the solar disk center in solar radii.}
  \label{F-lasco_wave1}
  \end{figure}

\begin{figure} 
  \centerline{\includegraphics[width=\textwidth]
   {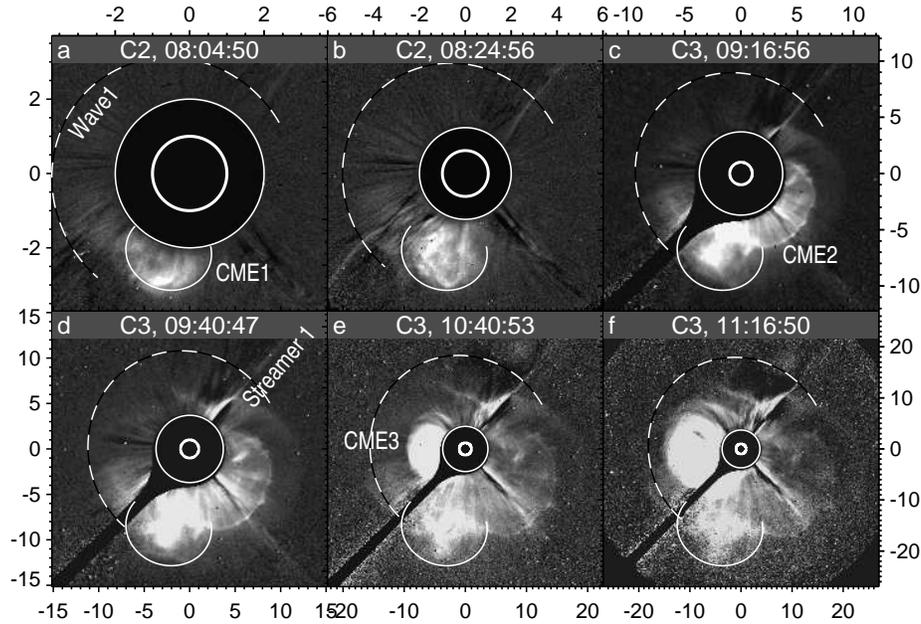}
  }
  \caption{CME1 in LASCO-C2 and C3 fixed-base ratio images resized
to compensate for the expansion of the CME. The solid white open
oval outlines the outer boundary of CME1. The dashed oval outlines
the traces of wave~1 (same as in Figure~\ref{F-lasco_wave1}). The
circles denote the solar limb and the inner boundaries of the fields
of view of the coronagraphs.}
  \label{F-lasco_cme1}
  \end{figure}

Figures~\ref{F-lasco_wave1} and \ref{F-lasco_cme1} allow one to
evaluate the quality of the measurements presented in
Figure~\ref{F-cme1_plot}. Figure~\ref{F-lasco_wave1} shows the
propagation of the faint wave~1 in LASCO images with a highly
enhanced contrast. All the images are progressively resized
following the measured kinematics to keep the visible size of the
dashed wave front constant. Propagation of wave~1 is solely revealed
by deflections of coronal rays (most likely, located not far from
the plane of the sky crossing the center of the Sun). The wave front
is most pronounced at position angles $\psi \approx
100^{\circ}-150^{\circ}$ being fainter at $\psi < 90^{\circ}$
(\textit{i.e.}, above the coronal hole\,---\,see
Figure~\ref{F-pre-event}d), and is additionally manifested in the
deflection of streamer~1. These properties correspond to an MHD
shock wave: the higher fast-mode speed above a coronal hole reduces
the Mach number, and therefore the shock front is not expected to be
pronounced there (\textit{cf}. \opencite{Grechnev2011_III}). The
wave speed in Figure~\ref{F-cme1_plot}b also supports its shock
regime, but dynamic radio spectra do not show a type II burst. It
seems that CME1 moves ahead of the associated wave front. Probably,
this visual effect is due to their different parallaxes,
\textit{i.e.}, because CME1 was considerably closer to SOHO than the
wave manifestations near the Sun's center plane.

Figure~\ref{F-lasco_cme1} shows LASCO-C2 and C3 images of the main
CME1 body (solid outline) resized according to the height-time plot
in Figure~\ref{F-cme1_plot}a. The dashed oval outlines wave~1 (same
as in Figure~\ref{F-lasco_wave1}). The structure of CME1 is not
identical in C2 and C3 images partly due to internal motions in the
CME and partly due to its changing visibility in the course of
expansion. The shape of the outlining oval is not obvious. Different
eccentricities of the ovals do not significantly change the
orientation of CME1 estimated in Section~\ref{S-ice-cream}; the
shape shown here is acceptable. Irrespective of the shape of the
oval, the heading structure of expanding CME1 remained south from
the ecliptic plane. Thus, its encounter with the Earth was unlikely
(the solar disk center corresponds to the Sun\,--\,Earth line). CME1
was able to produce, at most, a glancing blow on the Earth's
magnetosphere.

This analysis confirms the conclusion of Paper~I that the CME1 onset
was associated with the missed M1.2 flare at 07:29 in the east part
of AR~10501 and contrary to the idea of
\inlinecite{KumarManoUddin2011} about its association with episode
E2 at 07:41. Thus, eruption E2 was a confined one. Nevertheless,
this sharp impulsive eruption produced a shock wave.

\subsection{Shock 1 Produced by Confined Eruption at 07:41}

\begin{figure} 
  \centerline{\includegraphics[width=\textwidth]
   {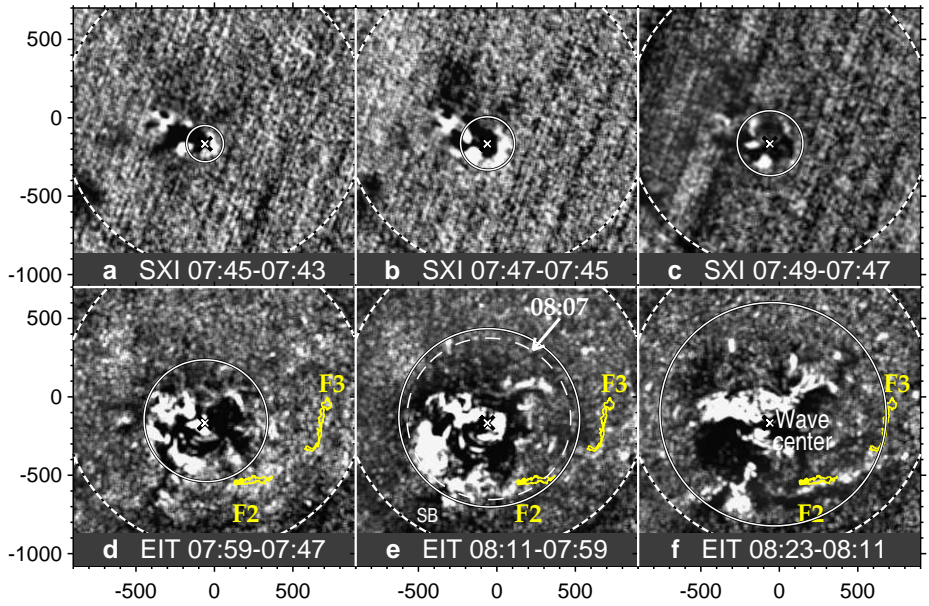}
  }
  \caption{Near-surface traces of shock~1 in GOES/SXI and
SOHO/EIT 195~\AA\ running difference-ratio images. The solid
ellipses calculated with $\delta = 2.55$, $t_0 = $~07:41:00 outline
the expanding shock front. The dashed ellipse in panel (e)
corresponds to 08:07. The yellow contours outline filaments F2 and
F3. The large dashed circles denote the solar limb.}
  \label{F-shock2_images}
  \end{figure}

Figure~\ref{F-shock2_images} presents traces of a shock wave
propagating near the solar surface in wide-band GOES/SXI images and
SOHO/EIT 195~\AA\ images produced with a lower imaging rate. The
SXI\_spectrum.mpg movie in the electronic version of the paper shows
the shock traces in GOES/SXI images (upper right corner) along with
the dynamic radio spectrum. The outline of the shock front in the
figure and the movie was calculated by using Equation
(\ref{E-pl_fit}) for propagation of a shock front along the
spherical solar surface with homogeneous distribution of plasma
parameters (the ellipses are intersections of the spheroidal wave
front with the spherical solar surface). We used $t_0 = $~07:41:00
and $\delta = 2.55$. The wave epicenter (slanted cross) is fixed at
slightly ahead of the visible edge of the ejection at $t_0$ (see
Paper~I). Traces of the expanding wave front are distinct in later
EIT images in the southeast to southwest directions. Most likely, a
fixed south brightening denoted `SB' in
Figure~\ref{F-shock2_images}e was due to eruption of CME1.

\begin{figure} 
  \centerline{\includegraphics[width=0.7\textwidth]
   {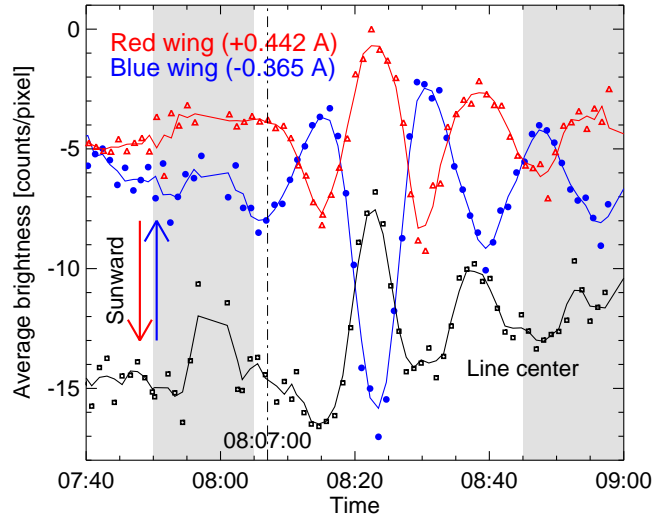}
  }
  \caption{Oscillations of filament F2
observed in the H$\alpha$ line center (black) and the red and blue
wings (KSO). The symbols present the measurements. The curves show
them smoothed over three neighbors. The shading marks the intervals
of cloudy weather. The arrows indicate the changes of brightness
corresponding to the sunward direction of the filament motion. The
vertical broken line marks a probable onset time (08:07) of the
anti-phase oscillations.}
  \label{F-winking_filament}
  \end{figure}

The near-surface portion of the shock front was distorted at a
large-scale inhomogeneity above the long filament channel traced by
filaments F2 and F3 (yellow in
Figures~\ref{F-shock2_images}d\,--\,\ref{F-shock2_images}f). The
shock front entered this enhanced-density region above filament F2
at about 08:07. The filament started to `wink' sequentially
appearing and disappearing in the red and blue wings of the
H$\alpha$ line. Figure~\ref{F-winking_filament} shows variations of
the average brightness of the whole filament F2 relative to its
close environment (photometry was made by an automated method).

Distinct anti-phase oscillations in the blue and red wings started
at about 08:07 (dash-dotted line) from the downward motion of the
filament pushed by the tilted shock front. The oscillations with a
period of 16 min probably reflect a self oscillation mode of the
whole filament but might be affected by a wave trail and arrival of
the second shock (discussed later) at about 08:15. A separate
analysis of the east, middle, and west portions of filament F2
showed that all the three parts oscillated in-phase with each other.

The fastest motion of the filament occurred at 08:23 in an upwards
direction, when it was darker in the blue wing than in the line
center. This indicates that its Doppler shift was larger than the
mid point between the blue wing and line center ($\approx
10$~km~s$^{-1}$). On the other hand, the absence of an overturn in
the blue-wing light curve in phase with the red wing near the valley
at 08:23 suggests that the Doppler shift did not exceed the blue
mid-wing wavelength ($\approx 20$~km~s$^{-1}$). Thus, the highest
line-of-sight velocity of the filament was $V_\mathrm{LOS} \approx
15$~km~s$^{-1}$ (\textit{cf.} \opencite{Tripathi2009}).

\begin{figure} 
  \centerline{\includegraphics[width=0.8\textwidth]
   {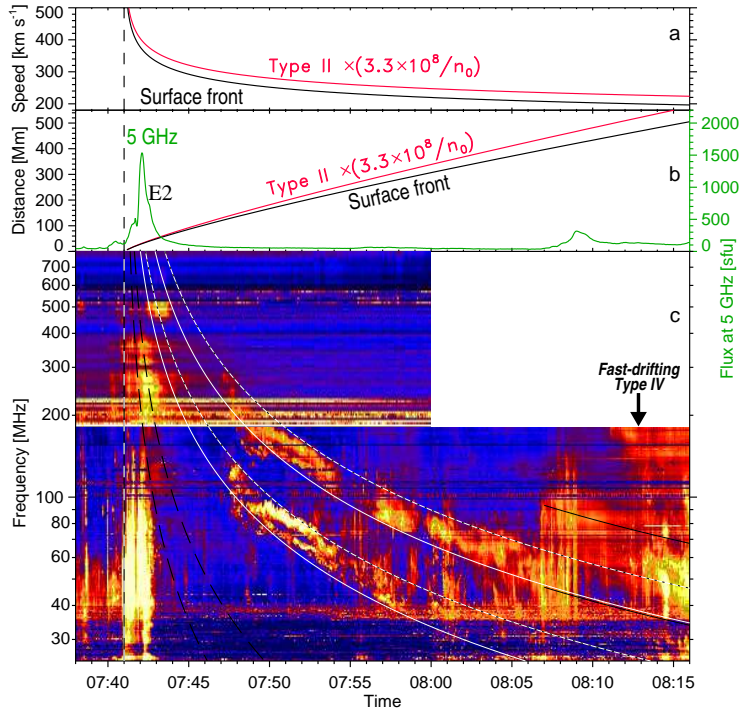}
  }
  \caption{Kinematics of shock~1 (a,b) in comparison with the
microwave burst E2 (green, b) and shock manifestations in the
dynamic spectrum of the first type II burst (c). The vertical dashed
line marks the shock onset time $t_0 = $~07:41:00. The two pairs of
white curves outline the split bands with the same $t_0$ and $\delta
= 2.60$. The black curves after 08:07 outline the N-like shift of
the bands. The steep black dashed curves outline possible signatures
of a quasi-parallel shock.}
  \label{F-shock2_spectrum}
  \end{figure}

The dynamic spectrum in Figure~\ref{F-shock2_spectrum}c composed
from the Culgoora (until 08:00), Learmonth, and San Vito data shows
a harmonic band-split type II burst. Its parameters are typical of
type II bursts associated with shock waves propagating upward in the
corona. The estimated shock speed was from 405 to 478 km~s$^{-1}$.

The outline for both pairs of the split bands was calculated as
described in Section~\ref{S-wave_expansion} with $\delta = 2.60$ and
the onset time $t_0 = $~07:41:00 (dashed vertical line), the same as
for the near-surface shock traces. The plots for the velocities and
distances \textit{vs.} time are shown in
Figures~\ref{F-shock2_spectrum}b and \ref{F-shock2_spectrum}c. Due
to the model dependence of estimates from radio spectra, the plots
for the type II tracer (red) are uncertain by a factor of $3.3
\times 10^8/n_0$, where $n_0$ is the actual plasma density at a
characteristic distance $h_0 \approx 100$~Mm. Near-surface shock
propagation and kinematics of the source of the type II burst
closely correspond to each other.

Comparison with near-surface shock traces in
Figure~\ref{F-shock2_images} shows that the type II burst started
when the shock front was located somewhere above regions 10501,
10503, and the bifurcation region. While the outline matches the
overall evolution of the drift rate, both actual bands deviate from
the outline like an inclined `S' by 07:54. The band splitting
disappears by 08:00. These properties disagree with a usual
interpretation of band spitting due to emissions from the downstream
and the upstream regions, implying instead emissions of split bands
from two extended coronal structures located close to each other
\cite{Grechnev2011_I}. The S-like deviation of the split bands and
their merger afterwards suggests that the shock front encountered a
high closed structure deflected by the shock.

At 08:07 the type II's bands underwent an N-like shift to higher
frequencies (black solid outline), suggesting that the shock front
entered an enhanced-density region. Figure~\ref{F-shock2_images}e
and `winking' filament F2 confirm that this really occurred at that
time. These facts along with the properties of the band splitting
indicate that the type II emission was most likely generated in a
nearly radial structure stressed by a quasi-perpendicular shock
(shock normal relative to the magnetic field). On the other hand,
fast-drifting features at about 07:42\,--\,07:45, which were
possibly harmonically related, hint at a possible much faster
quasi-parallel shock passage. The black dashed curves outline
possible harmonics.

\begin{figure} 
  \centerline{\includegraphics[width=0.6\textwidth]
   {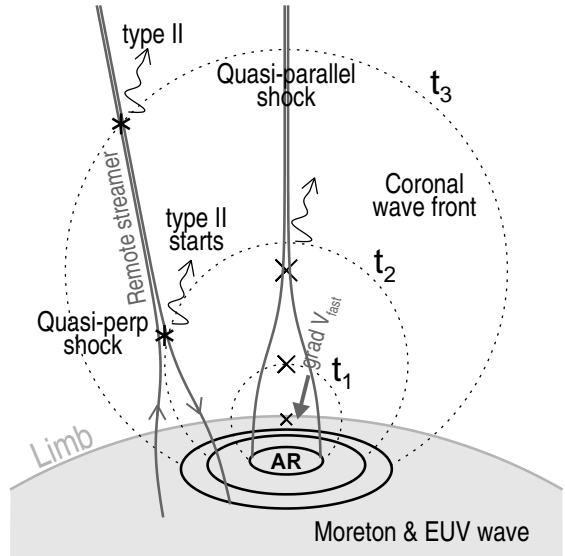}
  }
  \caption{Fast MHD shock wave excited by an impulsive eruption in
an active region (AR) and the appearance of type II emissions
excited by the quasi-perpendicular shock in a remote streamer and
by the quasi-parallel shock in the streamer above AR. The slanted
crosses denote the rising wave center at three consecutive times
$t_1$, $t_2$, and $t_3$.}
  \label{F-shock_cartoon}
  \end{figure}

A sketch in Figure~\ref{F-shock_cartoon} outlines our model of a
coronal wave excited in an active region (AR). The positions of the
wave front in the corona at three consecutive times $t_1$, $t_2$,
and $t_3$ are denoted by the dotted curves, and their corresponding
near-surface traces are shown with the solid ellipses. The arrow
$\mathbf{grad}\, V_\mathrm{fast}$ represents the conditions in the
low corona above the active region favoring the wave amplification
and formation of a discontinuity at $t_1$. The blast-like wave is
expelled from the AR core into regions of weaker magnetic fields.
The shock front crossing the current sheet inside a coronal streamer
excites a type II burst.

A wide-band type IV burst, which appeared after 08:11 at 180 MHz and
relatively rapidly drifted to lower frequencies, will be discussed
in Section~\ref{S-mosaic_spectrum}.

\subsection{CME2 at 08:49 and Shock 2}

\begin{figure} 
  \centerline{\includegraphics[width=0.8\textwidth]
   {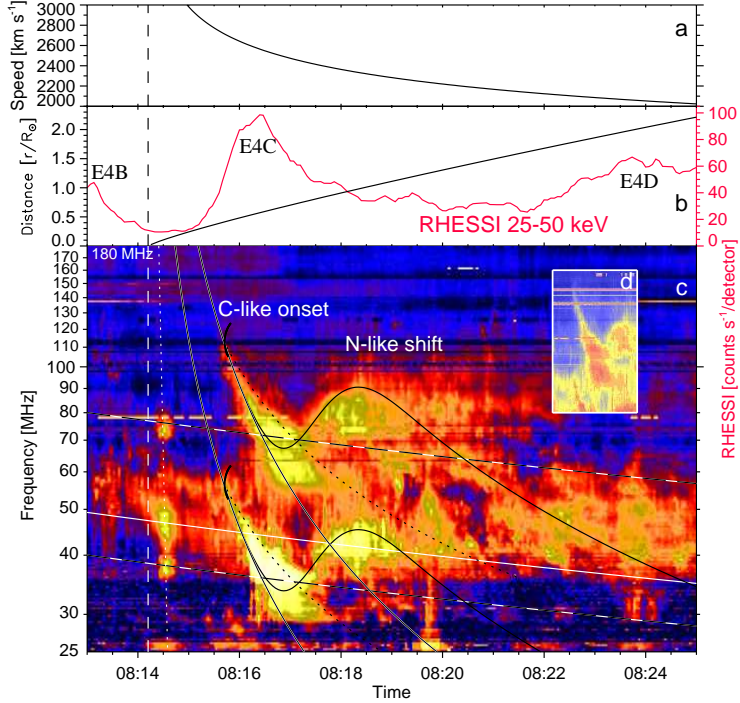}
  }
  \caption{Kinematics of shock~2 (a,b) and its manifestations in the dynamic
spectrum of the second type II burst (c). The red curve in panel (b)
is the RHESSI HXR flux. The inset (d) shows screen dump of the type
II onset in raw Learmonth file displayed by a standard viewer. The
almost vertical thin dotted line outlines the type III burst. The
nearly horizontal lines trace extensions of the first type II burst.
The remaining paired curves outline different harmonic components of
the second type II burst with $t_0 =
$~08:14:12 (vertical dashed line).}
  \label{F-shock3_spectrum}
  \end{figure}

To find a possible relationship between the expansion of CME2 and
radio signatures of the associated shock wave, the shock onset time
should be estimated. The highest accuracy of the estimation can be
achieved from the analysis of the radio spectrum in
Figure~\ref{F-shock3_spectrum}c, which was composed from the
Learmonth and San Vito data (its low-frequency part below 35 MHz is
suppressed due to interference).

The drift rate of the type II burst was atypically high and started,
in fact, from infinity. Its sharp C-like onset at about 08:15:35,
also visible in the inset (d), suggests a flatwise encounter of the
shock front with a nearly radial structure (see
\opencite{Grechnev2011_I}). Just after this encounter, the contact
region between the shock front and the streamer-like structure
bifurcated, and one emission source moved up, while another one
moved down thus producing the C-like feature.

Then both type II bands broadened considerably and underwent an
N-like shift to higher frequencies, while the initial bands possibly
continued. This behavior can be due to a portion of the shock front
entering into a denser region similar to the corresponding feature
of the first type II burst. The body of the second type II was
crossed by the bands of the first type II burst, whose drift rate
was much slower. They are outlined in
Figure~\ref{F-shock3_spectrum}c with a pair of dashed lines and a
white line (its corresponding fundamental band was below 25 MHz at
that time).

A probable onset time of the shock wave estimated from the drift
of the second type II burst falls within a valley between peaks
E4B and E4C in Figure~\ref{F-shock3_spectrum}b. The valley is due
to overlap of the decay of peak E4B and rise of peak E4C in the
total HXR emission. A probable onset time of peak E4C is marked by
a type III burst at 08:14:35 (crossed by the first type II). Type
III bursts are considered as prompt indicators of non-thermal
processes. By referring to this type III burst and extrapolating
its drift to its probable highest frequency of 2 GHz, we estimate
the shock~2 onset time $t_0 = $~08:14:12, which reconciles all its
considered manifestations. The drift of the type II burst can be
fit with an uncertainty of $t_0$ as large as $\pm 30$~s, while a
considerably wider uncertainty is allowable to fit expansion of
the outer halo component of CME2.

To outline the complex features of the type II burst, we adopt the
hypothesis of the shock front entering into a denser region. The
initial bands outlined with the black-on-white curves correspond to
$\delta = 2.65$. The black dotted curves outlining the
high-frequency boundaries of the broadened bands were calculated
with a considerably flatter density falloff $\delta = 2.1$. The
outline of the N-like feature was calculated by assuming a wide
Gaussian-shaped density enhancement in the way of the shock wave.
The complex structure of the type II burst and insufficient quality
of the dynamic spectrum does not allow us to understand the behavior
of the bands after 08:20.

\begin{figure} 
  \centerline{\includegraphics[width=\textwidth]
   {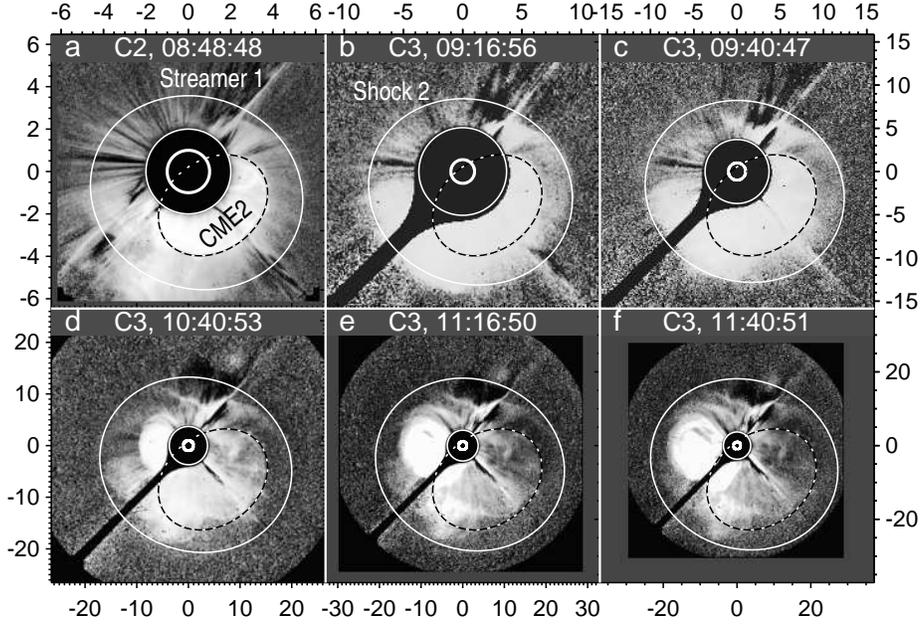}
  }
 \caption{Traces of shock~2 in resized LASCO-C2 and C3 running-ratio images.
The circles denote the solar limb and the inner boundaries of the
fields of view of the coronagraphs. The large white oval outlines
the outermost halo envelope of CME2. The dashed oval outlines the
outermost envelope of the arcade-like inner CME2 component (same
as in Figure~\ref{F-lasco_cme2}).}
  \label{F-lasco_cme_wave2}
  \end{figure}

The outer non-structured halo of CME2 outlined with the white oval
in Figure~\ref{F-lasco_cme_wave2} resembles traces of wave~1 in
Figure~\ref{F-lasco_wave1}. The shock-wave regime of the halo is
supported by the type II burst and features discussed later. We
therefore call the outer component `shock~2' and the inner one
`CME2'. Most likely, the eruption site of CME2 and source of
shock~2 were within a region limited by AR 10501, 10503, dimming
D2, and bifurcation region Rb (Figure~\ref{F-3cmes}d) rather close
to the solar disk center, which we adopt for simplicity as the
origin of the plots.

\begin{figure} 
  \centerline{\includegraphics[width=0.8\textwidth]
   {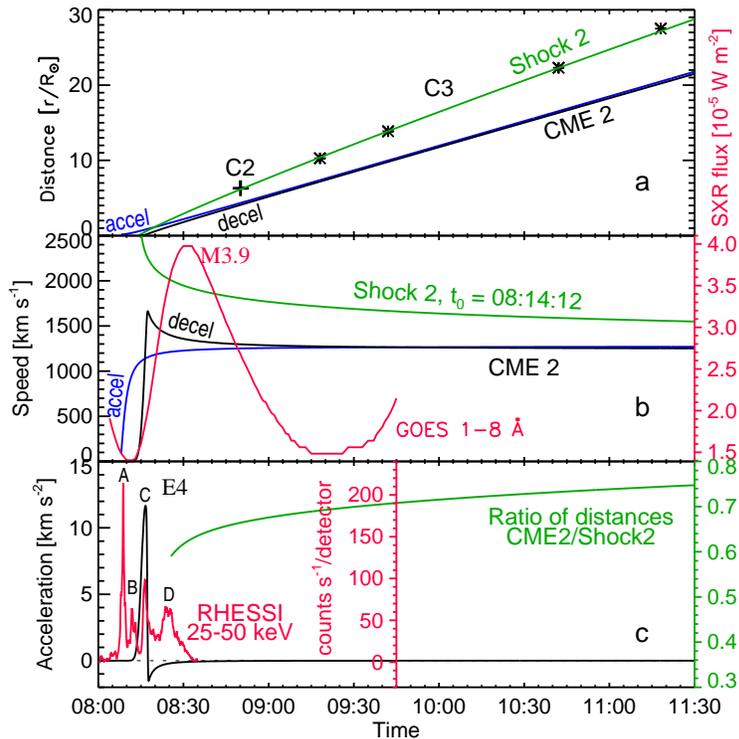}
  }
 \caption{Measurements of CME expansion for both
the wave (green) and arcade-like components (blue accelerating,
black decelerating). (a)~Height-time plot. The green curve fits the
shock wave. The symbols represent the measurements from the CME
catalog. (b)~Velocity-time plots in comparison with the GOES
1\,--\,8~\AA\ light curve (red). (c)~Acceleration of the CME along
with the HXR time profile (red) and the ratio of distances CME2 to
shock~2 (green).}
  \label{F-cme2_plot}
  \end{figure}

The green kinematical plot in Figure~\ref{F-cme2_plot}a calculated
by using Equation (\ref{E-pl_fit}) with the onset time of $t_0
=$~08:14:12 found from the dynamic spectrum agrees with the
measurements in the CME catalog of the fastest feature related to
CME2 (symbols). The white ovals outlining the halo envelope of CME2
in Figure~\ref{F-lasco_cme_wave2} (see also the CME2.mpg movie)
correspond to this curve. Deviations of streamer~1 ahead of shock~2
(which make shock~2 visible) are due to preceding wave~1. The
structure poleward from streamer~1 makes visible the streamer belt
deflected by shock~2. Concavity of the halo above the north pole
region is expected for a shock wave (Section~\ref{S-cmes_waves}).
These facts, as well as the high speed (green in
Figure~\ref{F-cme2_plot}b), strongly support the shock-wave nature
of the halo ahead of the main CME2 body (the dashed oval).

To coordinate expansion of the halo with the second type II burst,
we adjust the density model to bring the distances ($2.2R_\odot$)
and speeds (2000 km~s$^{-1}$) of the halo and the type II source
into coincidence at 08:25 (the ending time of
Figure~\ref{F-shock3_spectrum}). In fact, this assumption means a
spherical shock front propagating in an isotropic medium. Even with
this idealization, the difference between the speeds over the
plotted parts in Figures \ref{F-shock3_spectrum}a and
\ref{F-cme2_plot}b does not exceed 20\%. The corresponding reference
density $n_0 = 6.4\times 10^8$~cm$^{-3}$ is close to the Saito model
(see \opencite{Grechnev2011_I}). Figures~\ref{F-shock3_spectrum}a
and \ref{F-shock3_spectrum}b show the initial parts of the
kinematical plots for the shock~2 front calculated with this density
model. Comparison of the dynamic spectrum with the distance--time
plot in Figure~\ref{F-shock3_spectrum}b and images in
Figure~\ref{F-shock2_images} shows that the type II burst started at
a distance of $\approx 0.4R_\odot$ (08:15:35) from the source region
roughly corresponding to the position of filament F2, and the N-like
deviation started at $\approx 0.7R_\odot$ (08:16:50) roughly
corresponding to filament F3. The somewhat larger distance and the
gradual shape of the N-like deviation of type II-2 suggest a larger
height of its source relative to type II-1. This assumption is
consistent with the absence of the initial parts of the bands in
type II-2, which were split in type II-1; shock~2 probably developed
above the structure, from which these bands of type II-1 were
emitted.

\begin{figure} 
  \centerline{\includegraphics[width=\textwidth]
   {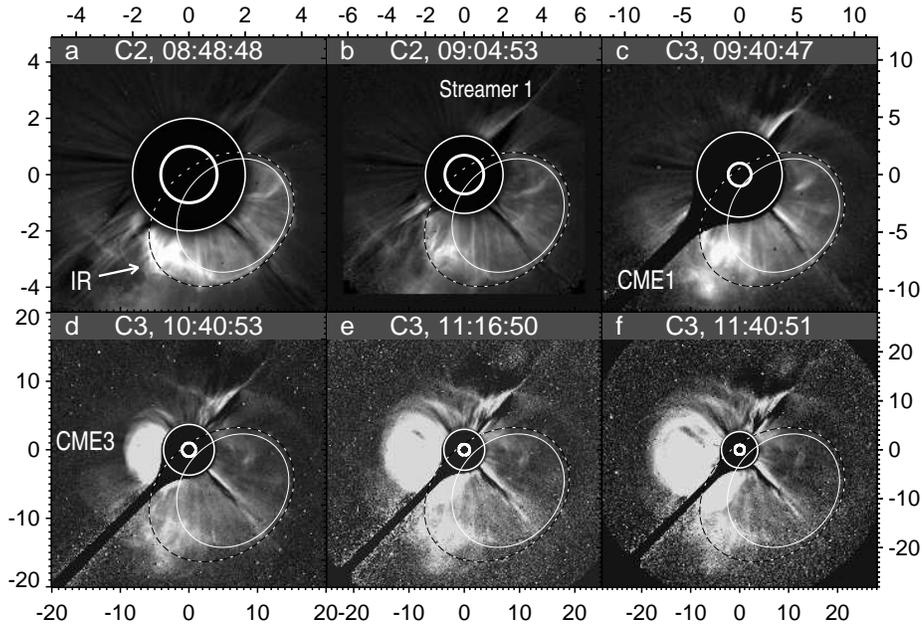}
  }
 \caption{CME2 in LASCO-C2 and C3 fixed-base ratio images resized
to compensate expansion of the CME. The white oval outlines the
spine of the main arcade-like structure. The dashed oval outlines
the outermost envelope of the arcade-like structure. The circles
denote the solar limb and the inner boundaries of the fields of
view of the coronagraphs.}
  \label{F-lasco_cme2}
  \end{figure}

The inner arcade-like component of CME2 had a pronounced spine
outlined in Figure~\ref{F-lasco_cme2} with the solid white oval.
The dashed oval outlines the outermost envelope of the inner
component including the intrusion region. Both ovals match the
expanding CME2. The height-time plot used in compensating its
expansion and plotting the ovals is shown in
Figure~\ref{F-cme2_plot}a.

Expansion of CME2 was nearly self-similar with minor deviations. To
keep the arcade spine within the white ovals, we slightly change
their parameters with time.
Figures~\ref{F-lasco_cme2}a\,--\,\ref{F-lasco_cme2}f reveal a
progressive displacement of the white oval southwest from the solar
disk center, \textit{i.e.}, from the Sun\,--\,Earth line. The main
leading part of CME2 is not expected to encounter the Earth. On the
other hand, the wide outermost part outlined with the dashed oval
increasingly covered the solar disk. These properties of CME2
indicate that its arcade-like part was directed southwest from the
Earth and, most likely, could only produce a glancing blow on the
Earth's magnetosphere. The intrusion region remained south of the
Earth.

\inlinecite{Gopal2005} estimated the onset time for the inner CME2
component as $\approx$~08:20 and its small acceleration. However,
our measurements outlining the whole CME2 show that its expansion
speed in the LASCO field of view was constant. LASCO images do not
allow us to understand whether CME2 accelerated or decelerated. We
compared plots for both kinematical types with X-ray light curves.
The latest possible onset time achievable for accelerating
kinematics corresponds to the blue curves in
Figures~\ref{F-cme2_plot}a and \ref{F-cme2_plot}b; later onset times
produce infinite results in Equations (\ref{E-self_sim_exp}) and
(\ref{E-onset_time}). The velocity starts to rise too early with
respect to the red SXR GOES plot. In this case it is difficult to
reconcile the velocity plots for the CME, shock, and the type II
burst.

By contrast, the decelerating type of kinematics (black curves)
provides acceptable results. The CME velocity in
Figure~\ref{F-cme2_plot}b starts to rise simultaneously with the SXR
emission. The decelerating self-similar part of the velocity plot
shows reasonable correspondence with the green shock wave plot. A
difficulty here is due to the fact that self-similar kinematics does
not describe the initial stage of rising acceleration. We have
described the impulsive acceleration stage with a Gaussian profile
(as we did in Paper~I; see also \opencite{Grechnev2011_I}), combined
the increasing velocity with the decreasing self-similar one, and
computed the distance and acceleration from the combined velocity.
The resultant impulsive acceleration up to $\approx 12$~km~s$^{-2}$
almost coincides with the HXR peak E4C, the deceleration peak of
about $-1.5$~km~s$^{-2}$ marks the onset of the self-similar stage,
and then acceleration decreases by the absolute value. Kinematical
plots with similar shapes and parameters have been previously
presented by \citeauthor{Temmer2008} (\citeyear{Temmer2008,
Temmer2010}) and \citeauthor{Grechnev2008} (\citeyear{Grechnev2008,
Grechnev2011_I}).

The green curve in Figure~\ref{F-cme2_plot}c presents the ratio of
distances CME2 to shock~2 from the eruption site (right $y$-axis).
The relative distance monotonically decreased for two reasons.
Firstly, CME2 moved nearly earthward, while the halo corresponded to
the lateral shock front, whose expansion was not facilitated by a
trailing piston. Thus, the lateral and especially rear shock was
closer to a freely propagating blast wave. Secondly, even the shock
front ahead of the CME2 tip decelerated and eventually must
transform to a pure bow shock.

\subsection{Overall Dynamic Radio Spectrum and an Extra Ejection}
 \label{S-mosaic_spectrum}

\begin{figure} 
  \centerline{\includegraphics[width=\textwidth]
   {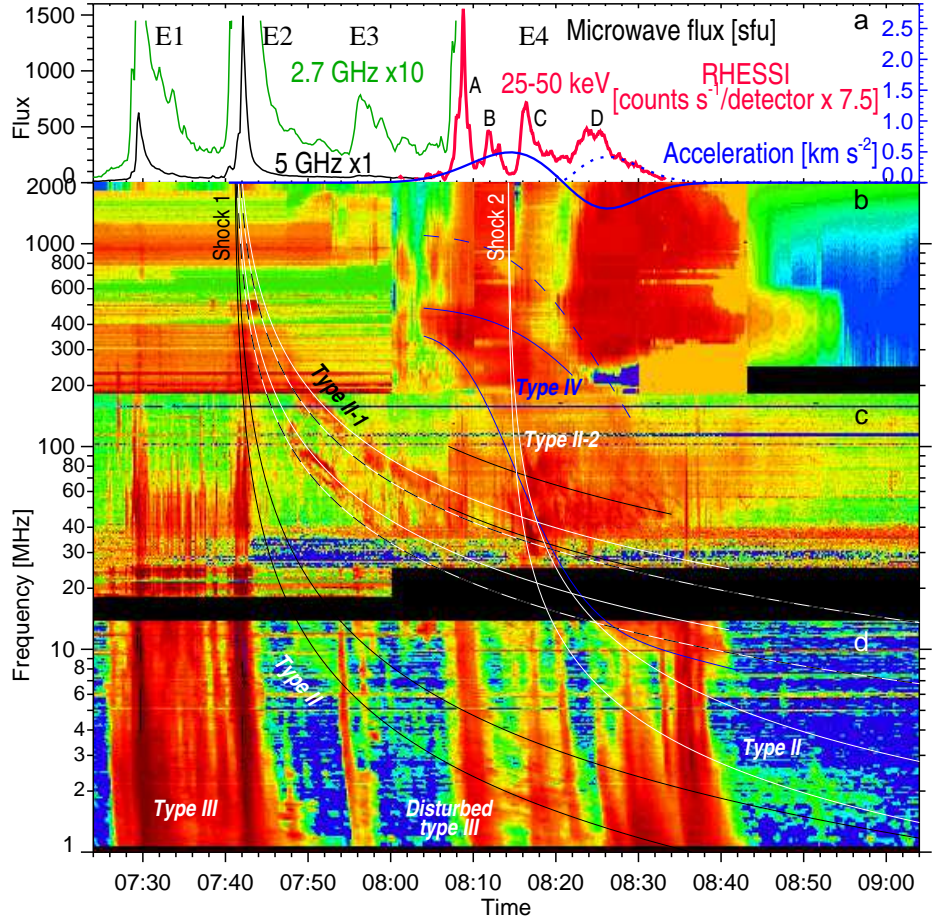}
  }
 \caption{Microwave (black and green) and HXR (red) time profiles (a)
and an overall dynamic spectrum composed as a mosaic from
observations of several instruments at decimeter, meter (b, c), and
decameter (d, \textit{Wind}/WAVES) wavelengths. The solid black,
white, and dashed black-white curves outline the type II bursts
(same as in Figures \ref{F-shock2_spectrum} and
\ref{F-shock3_spectrum}). The blue curves outline the fast-drifting
type IV burst. The leading blue low-frequency envelope of the type
IV burst was calculated from the acceleration presented with the
blue curve in panel (a). The dashed part of the acceleration plot
shows the absolute value of deceleration. The left $y$-axis in panel
(a) quantifies the microwave and HXR fluxes (see
Figures~\ref{F-timeprof}c and d). The right $y$-axis quantifies the
acceleration.}
  \label{F-mosaic_spectrum}
  \end{figure}

Figure~\ref{F-mosaic_spectrum} presents an overall picture of the
whole event including microwave and hard X-ray bursts E1\,--\,E4
(same as in Figures~\ref{F-timeprof}c and \ref{F-timeprof}d) and a
dynamic radio spectrum composed as a mosaic from pieces provided by
several observatories in different frequency ranges and time
intervals. The combined spectrum uses data from the Culgoora Solar
Observatory (18\,--\,1800 MHz) until 08:00 (b and c), Learmonth and
San Vito stations at 25\,--\,180 MHz (c), three parts form Bleien
Observatory (180\,--\,2000 MHz) at 08:00\,--\,08:43 (b), a set of
fixed-frequency records from San Vito to fill the gaps in panel (b),
and the \textit{Wind}/WAVES spectrum from the RAD2 receiver at
1\,--\,14 MHz.

The black and white curves of different line styles outline
signatures of the two shock waves discussed in the preceding
sections. The fast-drifting feature suggesting a quasi-parallel
shock~1 has a pronounced continuation at decameters after 07:48 (the
first pair of black lines) visible initially as a wide green band
and later traced by disturbed type III bursts during
08:13\,--\,08:22 (see, \textit{e.g.}, \opencite{Pohjolainen2008}).
The second type II burst also continues at decameters as a wide
green band between 1.5 and 3.5 MHz during 08:40\,--\,08:58 with
earlier indication of drifting features between the pair of the
white curves. Relating this drifting burst to interaction between
two CMEs proposed by \inlinecite{KumarManoUddin2011} is not
justified: this was a normal shock-associated type II burst. The
type II emission at decameters is presumably produced by the shock
front crossing a wide portion of the streamer belt with a relatively
wide range of densities that determines its wide frequency band.

The gap between the \textit{Wind}/WAVES spectrum and ground-based
observations hinders identification of the harmonic number for the
type II emissions at decameters. They are outlined assuming the
dominant fundamental emission, although a stronger harmonic emission
might be expected due to its weaker absorption. The alternative
outline is possible but requires a density falloff of $\delta
\approx 2.9$, which seems to be too steep at moderate latitudes.
Such an outline coordinated with the metric type II burst produces a
slightly higher drift rate at decameters than the observed one.
\inlinecite{CaneErickson2005} showed that the fundamental emission
at decameters sometimes dominates, which possibly justifies our
outline. Thus, we reproduce the drift rate of the decametric type
IIs, while identification of their harmonic structure remains an
open question.

Groups of type III bursts (especially clearly visible at decameters)
provide further support to our identification of the eruptions. A
dense type III group between 07:27 and 07:40 indicates the ongoing
escape of non-thermal electrons into open magnetic structures
probably associated with the CME1 liftoff, which started at E1. The
situation is drastically different after confined eruption E2, when
type IIIs rapidly terminate. Even the weak episode E3 produced a
clear type III response. A series of type IIIs marks the fourfold
event E4 suggesting a complex eruption, which has been partly
studied in Paper~I.

One more slowly drifting burst was reported as a type II by
observers in Bleien to occur at 08:04\,--\,08:33. However, its
evolution is opposite to the type IIs associated with shocks 1 and
2, and the bandwidth became quite broad. This burst is outlined with
the blue curves in Figure~\ref{F-mosaic_spectrum}. The solid curves
outline the suggested fundamental band, and the dashed curve
outlines a possible high-frequency envelope of the harmonic
emission. The trailing edge of this burst is difficult to recognize
and interpret.

The drift rate of this burst started from a near-zero value, which
excludes its relation to a wave. The large bandwidth suggests that
this was a type IV burst. It had an atypically high drift rate up
to very low frequency (but not exceptional---see, \textit{e.g.},
\opencite{Leblanc2000}). Relation of this burst to the body of
CME2 is unlikely due to the gradual acceleration up to the maximum
speed during 08:04\,--\,08:14 implied by the drift rate, whereas
CME2 sharply accelerated during E4C at about 08:16. Relating its
drift rate to the Saito or Newkirk density model has not resulted
in anything matching the observed CMEs.

\begin{figure} 
  \centerline{\includegraphics[width=\textwidth]
   {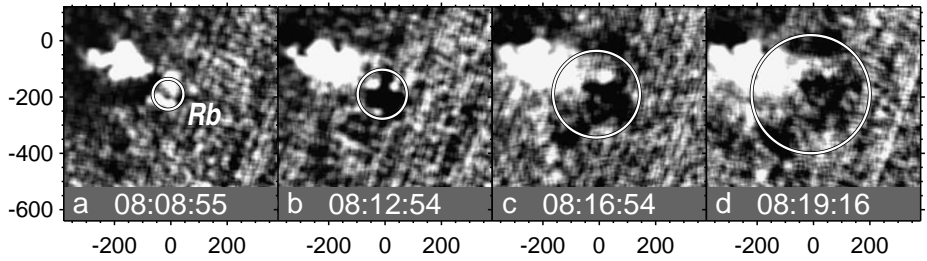}
  }
  \caption{Expansion visible in GOES/SXI running difference ratios
probably corresponding to the type IV burst. The radius of the
circle was calculated from acceleration in
Figure~\ref{F-mosaic_spectrum}a and exactly corresponds to the blue
outline of the leading low-frequency envelope of the type IV burst
in Figures~\ref{F-mosaic_spectrum}b\,--\,\ref{F-mosaic_spectrum}d.}
  \label{F-sxi_exp}
  \end{figure}

There is a different option. The lowest frequency of a radio burst
is determined by the plasma frequency $f_\mathrm{P} = 9 \times 10^3
n^{1/2}$ in an emitting volume. Assuming the frequency drift to be
due to the density decrease in an expanding spherical volume with
radius $r$, $n \propto r^{-3}$, we have adjusted acceleration (blue
in Figure~\ref{F-mosaic_spectrum}a) to match the low-frequency
envelope of the type IV burst. The initial density of $1.8 \times
10^{9}$~cm$^{-3}$ corresponds to 380 MHz. The spatial scale is
uncertain. With $r_0 = 30$~Mm corresponding to the bifurcation
region Rb, the initial part of the type IV burst's envelope
corresponds to the expanding motion visible in GOES/SXI images in
Figure~\ref{F-sxi_exp} (see also the SXI\_spectrum movie).
Manifestations of the expansion are not expected to be observed
later on, because the expanding feature moved away from the Sun. The
velocity of the latter motion cannot be estimated from the radio
spectrum.

The radial expansion of the ejection responsible for the type IV
burst accelerated up to $\approx 480$~m~s$^{-2}$ at about 08:14:22
(the radial speed at that time was $V_r \approx 180$~km~s$^{-1}$),
reached a maximum speed $V_{r\, \max} \approx 300$~km~s$^{-1}$, and
then decelerated to $V_{r\, \mathrm{final} }\approx
100$~km~s$^{-1}$. According to Paper~IV (Grechnev \textit{et al.},
in preparation), the average Sun\,--\,Earth transit speed of the
ICME responsible for the geomagnetic superstorm was $\overline{V}
\approx 865$~km~s$^{-1}$ (with an initial speed $V_0 \gsim
930$~km~s$^{-1}$). Thus, this ejection probably expanded within a
narrow cone with an angle of $2V_{r\, \mathrm{final}}/\overline{V} <
14^{\circ}$. Moving earthward almost exactly from the solar disk
center and expanding within such a narrow cone, this ejection should
appear in the LASCO-C2 field of view ($\ge 2R_{\odot}$) at a
distance $> 16R_{\odot}$ so that the Thomson-scattered light would
be meager. According to the estimates in Paper~I, the mass of this
ejection should be $\ll 5 \times 10^{15}$~g. The weak expansion and
low mass have made this CME invisible for LASCO.

\section{Discussion}
 \label{S-discussion}

\subsection{Shock Waves}
 \label{S-cmes_waves}

Analysis of the observations in the preceding section has revealed
a complex chain of CMEs and waves. Table~\ref{T-table2} summarizes
the results. The most noticeable fact is that the confined
eruption E2 undoubtedly produced a shock wave. Its presence is
confirmed by the type~II-1 burst, a detailed correspondence
between its drift and structure with the observed near-surface
propagation of the `EUV wave', the `winking' filament F2, and a
possible decametric type II burst due to the quasi-parallel shock.
All of these manifestations are quantitatively coordinated with
each other by the power-law description (\ref{E-pl_fit}) of an
impulsively excited shock wave quasi-freely propagating like a
decelerating blast wave.

 \begin{table}
 \caption{CMEs and waves revealed in the event.}
\label{T-table2}
 \begin{tabular}{cccl}
 \hline
\multicolumn{1}{c}{Time} & \multicolumn{1}{c}{Episode} &
\multicolumn{1}{c}{CME} & \multicolumn{1}{c}{Wave} \\
 \hline
07:29 & E1 & CME1 onset & Wave 1 \\
07:41 & E2 & No  & Shock 1 \\
08:14\,--\,08:16 & E4C & CME2 onset & Shock 2 \\
08:07\,--\,08:30 & E4A\,--\,E4D & Invisible CME  &  \\
 \hline
 \end{tabular}

 \end{table}

Paper~I has revealed that a portion of filament F1 was impulsively
heated between 07:39:59 and 07:41:27. The apparent speed of this
portion sharply reached $\approx 300$~km~s$^{-1}$ in the plane of
the sky, suggesting that its real speed along the filament leg was
$\approx 770$~km~s$^{-1}$ (at an angle of $\approx 23^{\circ}$),
which most likely produced considerable pressure pulse. This was
followed by an impulsive jet-like ejection with acceleration up to
2~km~s$^{-2}$ and a maximum speed of 450~km~s$^{-1}$ (both in the
plane of the sky). Each of these two impulsive phenomena could have
played a role of an impulsive piston; contributions from both are
possible. When the shock wave started, the related M3.2 flare only
began to gradually rise being unable to produce a significant
pressure pulse to excite the shock (\textit{cf.}
\opencite{Grechnev2011_I}). Eruption E2 had not produced any CME
which excludes the usually assumed bow-shock excitation by the outer
CME surface. This event presents a convincing pure case of shock
wave excitation by an impulsive eruption.

Similarly, shock~2 was excited during the early rise phase of the
E4C HXR burst in association with the onset of CME2. The velocity
and acceleration plots of CME2 (black in Figures~\ref{F-cme2_plot}b
and \ref{F-cme2_plot}c) demonstrate its impulsive-piston behavior,
while the propagation of shock~2 had the same decelerating pattern
as shock~1 (green in Figures~\ref{F-cme2_plot}a and
\ref{F-cme2_plot}b) described by Equation~(\ref{E-pl_fit}). The
shock-wave nature of this disturbance is confirmed by the
fast-drifting type II-2 burst traced up to decameters with its drift
rate and uncommon structural features described by the same
Equation~(\ref{E-pl_fit}), its super-Alfv{\' e}nic speed, and the
non-structured faint spheroidal halo in LASCO images
(Figure~\ref{F-lasco_cme_wave2}) both ahead of the arcade-like CME2
and well behind its rear part. There are additional features
expected for propagation of a shock wave.

\begin{figure} 
  \centerline{\includegraphics[width=\textwidth]
   {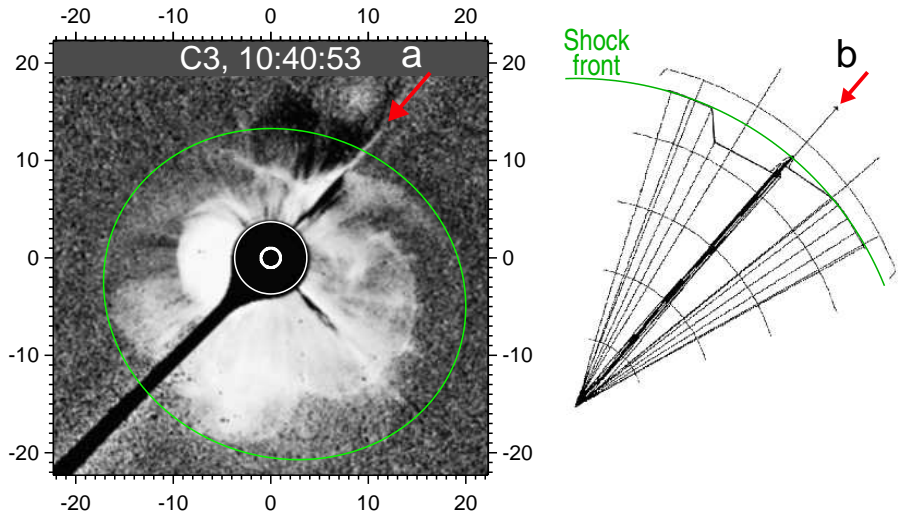}
  }
 \caption{Propagation of the shock front (green) along the streamer
belt: (a)~the outer halo envelope of CME2 observed by LASCO; (b)~the
calculated picture adopted from Uralova and Uralov (1994). The arrow
points at the streamer.}
  \label{F-streamer}
  \end{figure}

Figure~\ref{F-streamer} compares the halo envelope of CME2 observed
by LASCO-C3 with an expected distortion of the shock front in the
presence of the heliospheric current sheet (HCS) calculated by
\inlinecite{Uralova1994}. The red arrow in Figure~\ref{F-streamer}a
points at a coronal ray, which is a portion of the coronal streamer
belt aligned along the line of sight. This orientation makes it
distinctly visible. The streamer belt is the origin of the HCS.

\inlinecite{Uralova1994} addressed the propagation of a fast-mode
MHD shock wave along the HCS in the WKB approximation.
Figure~\ref{F-streamer}b presents Figure~5 of
\inlinecite{Uralova1994} rotated to correspond to the orientation in
Figure~\ref{F-streamer}a. The red arrow indicates the HCS inside a
radially diverging slow wind flow of enhanced density bounded by the
two long radial lines within $\pm 10^{\circ}$. A solar source of the
shock wave was considered apart from the HCS base on the solar
surface (not shown), which was located at the vertex of the ray
trajectories. The thick polygonal chain is the calculated shock
front far enough from the Sun (the polygonal shape was due to a
limited number of rays in the calculations). Its outermost portions
coincide with the green wave front calculated without the presence
of the HCS. A portion of the front in the close vicinity of the HCS
shown with the dashed arrow-like line represents the strongest
shock. It is due to the effect of regular energy accumulation in the
vicinity of the HCS. \inlinecite{Uralova1994} first suggested that a
small velocity component towards the HCS was able to initiate a
magnetic reconnection process accompanying a shock wave.

Comparison of Figures \ref{F-streamer}a and \ref{F-streamer}b shows
an overall qualitative similarity of distortions of the wave front
in the vicinity of the HCS that cause its concave shape. Unlike the
calculated picture, the real HCS in Figure~\ref{F-streamer}a is not
plane parallel to the line of sight. Its portion between the
streamer under the arrow and the dip nearly above the north pole has
been brought into view by the shock and corresponds to different
distances and position angles.

Shock~2 developed 33 min after the slower shock~1 at nearly the same
place in the plane of the sky and underwent the N-like shift of the
bands about 10 min after shock~1. This approach indicates that the
trailing shock~2 reached the leading shock~1 before its appearance
from behind the occulting disk of LASCO-C2. The two shocks should
combine into a single stronger one \cite{Grechnev2011_I}. Parameters
of shock~2 have unlikely changed significantly, because shock~1 was
much weaker. Due to probable coupling of the two shocks,
manifestations of shock~1 in LASCO images are not expected.

Our knowledge of wave~1 and related CME1 is poorer relative to
shocks 1 and 2. Its near-surface traces have not been detected,
neither was there a type II burst. On the other hand, traces of
wave~1 in LASCO images resembling a partial halo, the decelerating
kinematics also described by Equation (\ref{E-pl_fit}), and its
rather high speed of $> 850$~km~s$^{-1}$ up to at least
$10R_{\odot}$ indicate its shock-wave nature like shocks 1 and 2.
The absence of a type II burst and an `EUV wave' might be due to
different propagation conditions with its relatively low speed.

The widely presumed scenario of bow-shock excitation by the outer
surface of a CME is not confirmed. Ignition of a shock by a flare
pressure pulse is also unlikely \cite{Grechnev2011_I}. This
historically oldest scenario was based on an idea that the increase
of the plasma beta in flare loops up to $\beta \approx 1$ could
produce a significant disturbance. However,
\inlinecite{Grechnev2006} showed that the high-beta condition is a
normal situation in a flare. The plasma pressure in flare loops
increased due to chromospheric evaporation must be balanced by the
dynamic pressure of reconnection outflow coming from above. Even
with $\beta > 1$, the net effect is an increase of all sizes of a
flare loop as low as $\sqrt[4]{1+\beta}$, so that the expected
disturbance should be too small to produce a shock.

The major conclusion of this section related to the 20 November
superstorm is that the outer halo component of CME2 was most likely
a trace of a quasi-freely propagating shock wave and did not
indicate the earthwards direction of CME2.

\subsection{Consequences for a Problem of ``EIT Waves''}

Our analysis in Section~\ref{S-observations} touched the
long-standing challenging wave-like disturbances observed in EUV,
usually called ``EIT waves'' or ``EUV waves''. Debates over the
nature of these transients have lasted 15 years and do not appear to
have terminated so far (see, \textit{e.g.}, \citeauthor{Warmuth2010}
(\citeyear{Warmuth2010,Warmuth2011}) for a review). Their different
nature from the Moreton waves was prompted by their different
observed velocities and other properties seemingly inconsistent with
those of fast-mode MHD shock waves. A basic solution was initially
proposed by \inlinecite{Warmuth2001} and then developed by these
authors in several studies (\textit{e.g.}, \citeauthor{Warmuth2004a}
\citeyear{Warmuth2004a,Warmuth2004b,Warmuth2005}, and others). The
idea is that both kinds of phenomena are due to propagation of
decelerating fast-mode MHD shock waves. The Moreton waves are
usually observed at shorter distances, where the wave speed is
higher; EUV transients are observed at longer distances, where the
speeds of decelerating waves are lower. \citeauthor{Grechnev2011_I}
(\citeyear{Grechnev2011_I,Grechnev2011_III}) demonstrated that at
least two kinds of EUV transients visible as `EUV waves' did exist
and could be observed simultaneously. One kind of EUV transient is
due to plasma compression on top of a developing CME and by its
sides [basically consistent with the approach of
\citeauthor{Chen2002} (\citeyear{Chen2002,Chen2005})]. Near-surface
manifestations of such transients are of non-wave nature and remain
not far from an eruption site. The second kind of EUV transient
propagating over long distances is consistent with the initial
interpretation of the Moreton waves as lower skirts of coronal waves
proposed by \inlinecite{Uchida1968}. (Note in this respect the term
`coronal counterpart of a Moreton wave' used by some authors is
confusing.) The apparent discrepancies between properties of
propagating EUV transients and other shock signatures such as the
Moreton waves, type II bursts, and outer CME halos thus have a
simple explanation.

\inlinecite{Grechnev2011_I} showed that the most probable source of
an MHD shock wave is an impulsive eruption of a developing magnetic
flux rope. This is also consistent with the event in question. The
ends of an eruptive flux rope are fixed, while the velocity of the
eruption is highest in the direction of its expansion (often
non-radial, but mostly at a large angle with the solar surface).
Thus, an MHD disturbance excited by an impulsive eruption is
anisotropic, and the speed of its near-surface propagation is
considerably less than the upward one. For this reason, the
near-surface propagation velocity of an EUV transient is typically
much less than that of a type II source.

The fact that the Moreton waves are typically considerably faster
than EUV transients suggests that the Moreton waves are manifested
at stronger shocks than `EUV waves'. This circumstance is also
clear: to produce a Moreton wave, a shock wave has to penetrate to
relatively denser layers of the solar atmosphere that significantly
weakens the shock. By contrast, EUV signatures of a shock are
observed in higher coronal levels of lower density, so that
deceleration and damping of a shock does not prevent its observation
at much larger distances.

These circumstances show that reports on `winking filaments' driven
by `EIT waves', which were slower than type II burst sources, do not
contradict their excitation by shocks, as \inlinecite{Tripathi2009}
conjectured. A similar phenomenon considered in
Section~\ref{S-observations} present a confirmation. It should also
be noted here that the oscillating filament on 4 November 1997
reported by \inlinecite{Eto2002}, which was sometimes considered as
an argument against the shock-wave nature of `EIT waves', dealt with
an EUV transient poorly observed by EIT. By using the difference
ratios $-0.01 < I_\mathrm{wave} < 0.01$ (see
Section~\ref{S-kinematics}) of EIT images observed during this
event, one can detect faint but clear signatures of a propagating
disturbance at 06:13:54 at a much longer distance from the eruption
site than the authors found --- almost near a coronal hole at the
north pole.

\subsection{Orientations of the CMEs}
 \label{S-ice-cream}
To confirm and elaborate our preliminary conclusions about the
orientations of CME1 and CME2, now we try to employ a model which
allows one to estimate three-dimensional (3-D) geometric and
kinematical parameters of a CME observed by LASCO coronagraphs in
the plane of the sky. The so-called ice-cream cone model initially
proposed by \inlinecite{FisherMunro1984} considers a CME as a cone
with a vertex in the Sun's center. This model underwent several
elaborations. We use the model described by
\inlinecite{XueWangDou2005}. The model allows one to estimate the
radial velocity $|V|$ of a CME along its axis, the orientation of
the axis with respect to the Earth, and the angular width $\alpha$
of the CME cone. For our purposes it is convenient to express the
results provided by the model in an ecliptic longitude ($\lambda
>0$ west of the earthward direction) and latitude ($\phi >0$
north of the earthward direction).

To use the model of \inlinecite{XueWangDou2005}, an experimental
dependence is evaluated of the plane-of-sky velocity
$V_\mathrm{m}(\psi)$ of the CME envelope in LASCO images on the
azimuthal position angle $\psi$. Then a set of parameters
determining the orientation and axial speed of the CME is optimized
by using the least-squares fit of the measured set
$V_\mathrm{m}(\psi)$ to a calculated dependence $V_\mathrm{c}(\psi)$
(by minimizing the standard deviation $\sigma$). To expedite
adjustment of parameters in the optimization process, we employed a
genetic algorithm \cite{Mitchell1999}. Constraints on the fitting
parameters should be applied for implementation of this algorithm.
We used the following constraints: $1000 \leq |V| \leq 2000$
km~s$^{-1}$, $10^{\circ} \leq \alpha \leq 70^{\circ}$, and $\lambda$
and $\phi$ within $\pm 40^{\circ}$ relative to the axis passing from
the Sun's center through the CME source region.

\begin{figure} 
  \centerline{\includegraphics[width=0.85\textwidth]
   {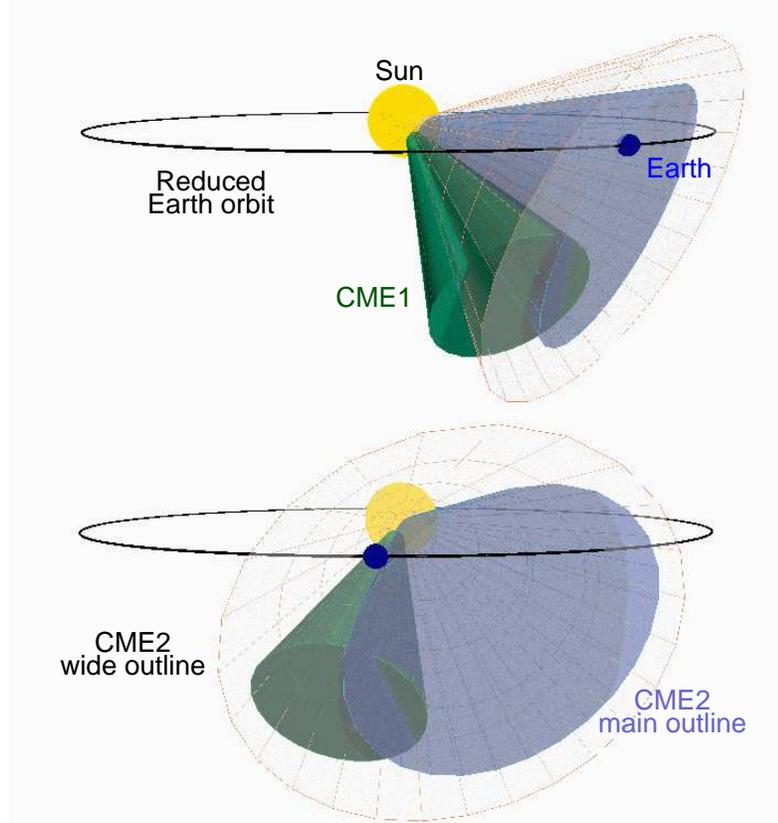}
  }
 \caption{Orientations of CME1 and CME2 estimated
by means of the ice-cream cone model. The top and bottom panels
represent different viewing directions.}
  \label{F-ice-cream}
  \end{figure}

3-D parameters of CME1 and CME2 were estimated from eight sets of
images observed with LASCO-C2 and C3. The contours of both main and
wide envelopes of CME2 in Figure~\ref{F-lasco_cme2} are well defined
with small uncertainties. This is not the case for CME1; estimations
of its 3-D parameters were additionally complicated by a narrower
range of position angles (see Figure~\ref{F-lasco_cme1}) which CME1
occupied, being far from the halo geometry. Therefore, extra
attempts were required to obtain better results for CME1. In these
attempts, we had to adjust velocity constraints for each iteration
by monitoring $\sigma$. Overall, the estimated parameters were
reasonably stable while input measurements were varied within the
limited ranges. The final results are listed in
Table~\ref{T-table3}. The corresponding sketch of the ice-cream
cones of CME1 and CME2 is shown in Figure~\ref{F-ice-cream} with
different viewing directions.

 \begin{table}
 \caption{Spatial parameters of CMEs estimated from the ice-cream cone model.}
\label{T-table3}
 \begin{tabular}{lcccccc}
 \hline
 \multicolumn{1}{c}{CME } & Time & Longitude$^{*}$ & Latitude$^{*}$ & Span$^{*}$ & Speed $|V|^{*}$ & Deviation \\

 \multicolumn{1}{c}{No.} & interval & $\lambda \ [^{\circ}]$ & $\phi \ [^{\circ}]$ & $\alpha \ [^{\circ}]$ &
 [km s$^{-1}]$ & $\sigma$ [km s$^{-1}$] \\

 \hline
1 & 08:05--11:41  & $-8 \pm 0.7$ & $-26 \pm 1.8 $ & $28 \pm 2.0 $  & $1950 \pm 24$  & 8.1--13.5 \\
2 Main & 08:49--12:17  & $17 \pm 1.4 $ & $-16 \pm 1.2 $ & $50 \pm 2.4 $  & $1778 \pm 9$ & 1.0--1.8 \\
2 Wide & 08:49--12:17  & $13 \pm 1.4 $ & $-18 \pm 1.7 $ & $66 \pm 2.5 $  & $1718 \pm 55$ & 3.2--4.9 \\
 \hline
 \end{tabular}
$^{*}$Average and range of estimates from different images in the
interval specified in column 2.
 \end{table}

Table~\ref{T-table3} and Figure~\ref{F-ice-cream} confirm our
preliminary conclusion that both CME1 and CME2 were not directed
exactly earthwards. Each of the CMEs propagated mainly southward
from the ecliptic plane, being only able to produce a glancing blow
on the Earth's magnetosphere. Ongoing expansion of an ICME suggests
that the magnetic fields at its flanks were significantly weaker
than at its nose. Due to magnetic flux conservation, the magnetic
field strength at a fixed position of a self-similarly expanding
ICME is inversely proportional to its instantaneous size squared
(and the speed decreases linearly). For example, if an ICME flank
hits the Earth at a distance of $1/\sqrt{2}$ of the heliocentric
distance of the ICME nose, then the magnetic field at the flank
should be reduced by a factor of 2 with respect to the central
encounter. To our knowledge, the total magnetic field strength
$|\mathbf{B}| \approx 56$~nT in the 20 November 2003 magnetic cloud
was close to a record one. Still stronger fields were only observed
in November 2001: on 6th, $|\mathbf{B}| \approx 66$~nT, and on 24th,
$|\mathbf{B}| \approx 57$~nT (A.~Belov, 2012, private
communication). If the encounter of the 20 November 2003 MC with the
Earth were a non-central encounter, one would have observed
significantly stronger magnetic field; which is unlikely.

Thus, direct responsibility for the superstorm of magnetic
structures of CME1 or CME2 appears to be doubtful. On the other
hand, the mutual lateral pressure of CME1 and CME2 should
considerably affect their expansion as well as any structures
between them including the hypothetical invisible CME. This
circumstance hints at possible causes of its weak expansion.

\subsection{Eruption near the Solar Disk Center}

Now we have sufficient information to assume what could have
occurred near the solar disk center between 08:07 and 08:17. Paper~I
has established that the eruptive filament F1, which lifted off at
an angle of $\approx 60^{\circ}$ to the solar surface, at about
08:07 collided with a topological discontinuity and bifurcated. The
major mass of the filament moved nearly along the solar surface
afterwards and had not left the Sun. At the same time and place, a
nearly spherical structure developed and erupted with an initial
speed of motion away from the Sun of $\gsim 930$~km~s$^{-1}$. Its
very slow expansion almost exactly from the solar disk center (the
established radial expansion speed of $\approx 100$~km~s$^{-1}$) and
the earthward orientation have made it invisible for the LASCO
coronagraphs. The only reasonable cause of its development was the
anomalous collision of the eruptive filament F1 with a magnetic
obstacle.

Most likely, one more product of this collision was the development
of the coreless arcade-like CME2. Magnetic fields in a pre-eruption
arcade are nearly potential (rot$\mathbf{B} \approx 0$), and
therefore the arcade was unlikely to erupt by itself. Thus, CME2 was
probably forced to erupt being hit from below. Its onset time of
about 08:15 indicates that its probable cause was also the eruptive
filament F1, whose active role was established in Paper~I. This
assumption is supported by decelerating kinematics of CME2 (see
Figure~\ref{F-cme2_plot}b). The observations lead therefore to the
following picture. The \textit{magnetic flux rope} developed
\textit{from filament F1} and moved southwest with an initial angle
of $\approx 60^{\circ}$ to the solar surface ($\approx 23^{\circ}$
to the line of sight). When passing through the topological
discontinuity near the solar disk center at a height of $\approx
100$~Mm, the eruptive filament (flux rope) caused an expansion of
the arcade above it (in a normal case, the arcade would be a CME
frontal structure), but failed to become its core. Instead, the
filament disintegrated into two parts, one of which remained on the
Sun, and the other one erupted as a `core' (invisible CME), but
apart from CME2. The initial velocity of the invisible CME $\gsim
930$ km~s$^{-1}$ is comparable with the initial speed of CME2
($\approx 1700$ km~s$^{-1}$), confirming their association and the
assumption that the eruption of CME2 was forced by the eruptive
filament F1. Development of shock~2 at 08:14:12 was most likely
related to this violent episode.

Separation of CME2 into the `coreless CME' and `CMEless core'
(without the frontal structure) hints at a more complex relation
between the CME parts than traditionally assumed. The core might be
an active CME component responsible for its initiation and initial
propagation, and the frontal structure might be a passive envelope
arcade whose expansion is driven from inside. Note that the
appearance of CME3 in Figure~\ref{F-3cmes}c supports this
assumption: its core was pronouncedly twisted suggesting active
motions followed by a kink instability, while the outer structures
of CME3 consisted of steadily expanding closed long loops rooted on
the Sun. After relaxation of the core, the whole CME expanded
self-similarly. The difference between the loops in the structures
of CME2 and CME3 was due to their orientations. Unlike CME2, in
which the planes of the arcade loops were close to the line of
sight, the planes of the frontal loops of CME3 were close to the
plane of the sky.

The joint analysis of the dynamic radio spectrum and GOES/SXI images
has shown that HXR peak E4D (the last one whose association was not
revealed) corresponds to deceleration of the invisible CME. As
discussed in Paper~I, considerations and results of several
researchers converge to the conclusion that HXR and microwave bursts
presented a flare manifestation of magnetic reconnection responsible
for acceleration of a developing flux rope, when the propelling
toroidal force developed. Similarly, the deceleration reflected by
the HXR peak E4D might be a response to another reconnection
process. This process possibly destroyed magnetic structures
providing the toroidal force so that only retarding magnetic tension
responsible for deceleration persisted, and then the eruption
probably disconnected completely, thus entering a free expansion
stage. This speculation implies that HXR and microwave bursts
indicate both acceleration and deceleration of CMEs, and that the
self-similar expansion began, when the flare bursts ceased.

\section{Conclusions}
 \label{S-conclusion}

Our detailed analysis of the complex solar eruptive event carried
out in this paper and Paper~I has led to a number of results, which
are not only important in pursuing causes of the 20 November 2003
geomagnetic superstorm, but also are promising for better general
understanding of solar eruptions, CMEs, related shock waves, and
their various manifestations. In particular, identification of an
outer halo CME component with a shock trace promises better
estimates of orientation and velocity of CMEs and higher accuracy in
predicting the arrival time of a corresponding ICME.

The shock waves revealed in this event provide further support for
the concept of early impulsive-piston shock excitation by an
eruptive structure proposed by \inlinecite{Grechnev2011_I}. A shock
wave excited by a confined eruption at 07:41 presents a notable
example confirming this scenario. On the other hand, the widely
presumed bow-shock excitation scenario at the outer surface of a CME
is not confirmed. Ignition of a shock by a flare pressure pulse is
also unlikely.

Magnetic structures of neither CME1 nor CME2 appear to be
appropriate candidates for the sources of the superstorm for the
following reasons.

\begin{itemize}

\item
 CME1 erupted, most likely, at about 07:29 from the east part of AR~10501,
where the helicity was excessively negative. CME1 was not
earth-directed.

\item
 The outer halo of CME2 was probably due to a spheroidal shock front
and did not indicate the earthward direction of magnetic structures
of CME2.

\item
 Expansion of CME1 and CME2 close to each other probably caused
their mutual compression, but there were no signs of reconnection
between their magnetic structures.

\item
 CME1 and CME2 were directed southward from the ecliptic
plane, oblique with respect to the Sun\,--\,Earth line, being only
able to produce a glancing blow on the Earth's magnetosphere with a
reduced geomagnetic effect.

\end{itemize}

These circumstances disfavor the idea of \inlinecite{Chandra2010}
about a positive-helicity eruption from AR~10501. The suggestions of
\inlinecite{KumarManoUddin2011} and \inlinecite{Marubashi2012}
related to the causes of the 20 November 2003 superstorm lose their
basis. On the other hand, GOES/SXI and radio observations provide
further support to the presumed additional CME which erupted close
to the solar disk center. Its estimated characteristics confirm the
assumption made in Paper~I that its weak expansion within a narrow
cone of $< 14^{\circ}$ could make it invisible for LASCO and
preserve its very strong magnetic field due to magnetic flux
conservation.

Our study demonstrates that even a case study of a single event
can supply rich information about solar eruptions, associated
phenomena, and their consequences. The major condition of success
was a combined analysis of multi-spectral data. It has been
recognized that significant suggestions and milestones are
provided by bursts generated by accelerated electrons. They are
observed as flare bursts in hard and soft X-rays and microwaves as
well as drifting radio bursts at longer radio waves. Our results
emphasize particularly the following.

\begin{itemize}

\item
 Type III bursts are well-known signatures of non-thermal
electrons. Their appearance can be indicative of acceleration
processes occurring during eruptive episodes. In particular, our
event demonstrated dense trains of type III bursts accompanying
the CME lift-off.

\item
 The concept of predominant excitation of type II bursts by
decelerating quasi-perpendicular shocks in remote streamers allowed
us to reconcile their various features with other signatures of
propagating shock waves. In particular, this concept accounts for
the delay of the type II onset time relative to HXR and microwave
flare bursts and the relatively low starting frequencies of type II
bursts. The latter becomes clear if one considers the tilted shock
front excited at a height of $\approx 100$~Mm to encounter a remote
streamer at some distance from the eruption site.

\item
 The type IV burst discussed here was possibly a moving type IV, but
we cannot confirm this possibility due to the absence of meter-wave
imaging observations. The approach used here promises diagnostics of
developing CMEs from type IV bursts with relatively fast drift.

\end{itemize}

In summary, the combined analysis of the multi-spectral observations
carried out in Paper~I and this paper makes it possible to construct
a consistent picture of several observational facts and suggestions,
some of which seemed to have been questionable. The outlined
scenario accounts for most of these circumstances. Unanswered
questions still remain, however. It is unclear what occurred in the
magnetic structures of the eruptive filament in the bifurcation
region, how the `CMEless core' was formed, and how to reconcile the
right-handed magnetic cloud with the left-handed pre-eruption
structure. These issues will be addressed in Paper~III. One more
question is specifically what kind of structure reached the Earth on
November 20 and produced the superstorm. This will be a subject of
Paper~IV.

\begin{acks}

We thank Viktoria Kurt for the CORONAS-F/SONG data,
L.~Kash\-ap\-ova and S.~Kalashnikov for the assistance in data
processing, and I.~Kuzmenko for useful discussions. We are
grateful to an anonymous reviewer for useful remarks. We thank the
instrumental teams of the Kanzelh{\"o}he Solar Observatory; MDI,
EIT, and LASCO on SOHO (ESA \& NASA project); the USAF RSTN Radio
Solar Telescope Network; RHESSI; and the GOES satellites for the
data used here. We thank the team maintaining the CME Catalog at
the CDAW Data Center by NASA and the Catholic University of
America in cooperation with the Naval Research Laboratory. This
study was supported by the Russian Foundation of Basic Research
under grants 11-02-00757, 11-02-01079, 12-02-00008, 12-02-92692,
and 12-02-00037, The Ministry of education and science of Russian
Federation, projects 8407 and 14.518.11.7047. The research was
also partly supported by the European Commission's Seventh
Framework Programme (FP7/2007-2013) under the grant agreement
eHeroes (project No. 284461), \url{www.eheroes.eu}.

\end{acks}

\end{article}

\end{document}